\documentclass[reprint, author-year]{JASA}

\usepackage{subfigure}
\usepackage{tikz}
\usetikzlibrary{arrows.meta,positioning,shapes.geometric,fit,calc,backgrounds,decorations.pathmorphing}

\graphicspath{{Images/}}
\newcommand{\pnl}[4]{\begin{minipage}[t]{#1}\raggedright{\sffamily\fontsize{8}{9.6}\selectfont\textbf{#2}\enspace #3}\\[0.6mm]%
  \includegraphics{#4}\end{minipage}}
\tikzset{
  fcbox/.style={rectangle, rounded corners=2pt, draw=black!70, thick,
    align=center, inner sep=5pt, minimum height=8mm, fill=black!3, font=\small},
  fcio/.style={rectangle, draw=black!70, thick, align=center, inner sep=5pt,
    minimum height=8mm, fill=black!8, font=\small},
  fcflow/.style={-{Latex[length=2mm]}, thick},
  fcadj/.style={-{Latex[length=2mm]}, thick, dashed, draw=blue!55!black},
  fclabel/.style={font=\scriptsize\itshape, text=black!60},
}

\usepackage{hyperref}
\begin{document}

\title[DART: Differentiable Acoustic Ray Tracing]{DART: Differentiable Acoustic Ray Tracing with Gaussian Beams for Underwater Sound Propagation and Inversion}

\author{Indu Kant Deo}
\email{indukant@student.ubc.ca}
\affiliation{Department of Mechanical Engineering, The University of British Columbia, Vancouver, British Columbia V6T 1Z4, Canada}

\author{Tushar Kant Deo}
\email{tusharkantdeo@gmail.com}
\affiliation{Independent researcher, Bangalore, Karnataka 560103, India}

\author{Akash Venkateshwaran}
\email{akashv22@student.ubc.ca}
\author{Rajeev K. Jaiman}
\email{rjaiman@mech.ubc.ca}
\affiliation{Department of Mechanical Engineering, The University of British Columbia, Vancouver, British Columbia V6T 1Z4, Canada}


\begin{abstract}
Underwater sound propagation depends on sound-speed structure, bathymetry, and seabed properties. Reliable derivatives with respect to these environmental inputs are therefore needed for acoustic inversion and uncertainty assessment.
This study presents DART (Differentiable Acoustic Ray Tracing), a differentiable geometric hat- and Gaussian-beam solver for two- and three-dimensional underwater sound propagation implemented in JAX on a graphics processing unit. 
The formulation computes sensitivities consistent with the forward model by differentiating ray paths, event locations, interface jumps, beam spreading, and receiver contributions, while a sparse adjoint and trajectory recomputation reduce memory use during reverse-mode differentiation.
At matched inputs, root-mean-square transmission-loss differences are 0.51 and 1.16 dB from BELLHOP for the Munk and Dickins benchmarks, respectively, and 0.09–0.26 dB from BELLHOP3D across three cases.
Forward- and reverse-mode derivatives agree closely, and independent centred finite-difference checks verify the selected sensitivities.
A gradient with respect to 25,921 bathymetry nodes costs 5.9 forward solves, while larger kernels are computed by dividing the beam fan into azimuthal blocks.
Environmental kernels identify the seabed regions and sound-speed depths that influence the acoustic field. Within the tested linear range, these sensitivities predict acoustic responses to environmental changes. Synthetic inversion demonstrates environmental parameter estimation and quantifies local posterior uncertainty. These capabilities can support acoustic digital twins by updating environmental estimates from observations and assessing how changing ocean conditions affect predicted noise exposure.
\end{abstract}
\maketitle
\section{\label{sec:1} Introduction}

Underwater radiated noise from marine vessels affects marine animals that rely on sound for communication, navigation, and prey detection~\cite{duarte2021soundscape,erbe2019effects,richardson1995marine, Deo_2025_thesis}. Growing maritime activity has contributed to increasing underwater noise in several ocean regions~\cite{erbe2019effects, venkateshwaran2024multi,venkateshwaran2025mute}. In response, the International Maritime Organization (IMO) has issued voluntary guidelines encouraging noise reduction through ship design, maintenance, and operational measures~\cite{imo2024urn}. Assessing these measures requires predicting how vessel noise propagates through the ocean. Because the received field depends on both the source and the environment, changing ocean conditions can alter acoustic exposure even when the source remains unchanged.

Sound-speed structure, bathymetry, and seabed properties influence refraction, reflection, and geometric spreading~\cite{jensen2011computational}. Established propagation methods include ray and Gaussian-beam tracing~\cite{porter1987gaussian,porter2019beam,porter2011bellhop}, parabolic-equation methods~\cite{collins1989applications,tappert1977parabolic}, and normal-mode formulations~\cite{porter1984normal}. Their computational cost depends on frequency, propagation range, environmental complexity, and numerical resolution. Even when a single prediction is affordable, repeated solves across different environments can make uncertainty quantification and inversion computationally demanding. We use Gaussian-beam propagation to model refraction, beam spreading, and boundary interactions.

Environmental inversion uses acoustic observations to estimate properties of the ocean and seabed~\cite{baggeroer1993overview,dosso2002quantifying}. Local sensitivity analysis uses derivatives to identify which environmental parameters most strongly affect the predicted field. Gradient-based inversion uses these derivatives to update parameter estimates and reduce the mismatch between predictions and observations. For gridded bathymetry, estimating sensitivities by perturbing each node separately requires a number of propagation solves that grows with the grid size~\cite{griewank2008evaluating}. The derivatives must also be consistent with the numerical model used to predict the field. Agreement between forward predictions alone does not establish agreement between their derivatives.

Adjoint methods and Fr\'echet kernels are established tools in geophysics and seismic tomography~\cite{plessix2006review,tromp2005seismic,dahlen2000frechet}. In underwater acoustics, related work includes travel-time sensitivity kernels for ocean tomography~\cite{munk1995ocean,skarsoulis2004travel} and adjoint methods for acoustic inversion~\cite{hursky2004adjoint}. Automatic differentiation has been applied to full-wave simulation~\cite{stanziola2023jwave, deo2025differentiable}, higher-order parabolic-equation modelling for sound-speed profile estimation~\cite{lytaev2024automatically}, and normal-mode sensitivity analysis~\cite{vardi2026sensitivity}. Adjoint methods have also been used to differentiate curved rays in optics~\cite{teh2022adjoint}, while learned acoustic surrogates approximate the mapping from environmental inputs to acoustic fields~\cite{deo2022POF, deo2023_3dCOnvPOF, DEO2026JCP, mallik2022predicting,deo2024continual,deo2025data}. For beam models, parallel C++ and CUDA ports of BELLHOP and BELLHOP3D accelerate forward calculations~\cite{pisha2023bellhopcuda}. \citet{chitre2023differentiable} presented differentiable propagation models, including a two-and-a-half-dimensional Gaussian-beam tracer that uses forward-mode differentiation, and demonstrated gradient-based inverse applications.

Differentiating a beam solver requires accounting for how the trajectory and its contributions at receivers change with the environment. A perturbation can shift a boundary contact or a sound-speed knot crossing, changing where the solver applies a state update. Caustic phase changes and decisions about which receivers lie within a beam's support also affect the assembled field. The derivative calculation must therefore follow the event locations and state updates used in the forward discretization. Changes in event order, the seabed facet encountered, or beam support make the computed field piecewise smooth. We use finite perturbations to test when local derivatives can predict the effects of environmental changes.

This study presents DART, a differentiable JAX solver~\cite{jax2018github} for the geometric hat and Gaussian beams of \citet{porter2019beam}. We follow the differentiable programming approach used in compressible flow~\cite{bezgin2023jaxfluids}. The solver supports two- and three-dimensional propagation, with out-of-plane refraction and coupled $2\times2$ dynamic matrices in three dimensions. It combines batched GPU execution with reverse-mode differentiation of the discrete beam calculation. A custom adjoint for sparse receiver accumulation and trajectory recomputation limit reverse-mode memory. We verify the fields against BELLHOP and BELLHOP3D at matched inputs and check selected derivatives using forward mode, reverse mode, and centred differences. In the cases tested, a single-pass gradient of a scalar objective costs the equivalent of 3.3--5.9 forward solves. The largest three-dimensional kernels are computed in azimuthal blocks to fit device memory. Environmental sensitivity kernels and synthetic inversion show how the derivatives can be used to interpret field changes and estimate environmental parameters. Sections~\ref{sec:2} and~\ref{sec:3} describe the formulation and implementation. Section~\ref{sec:5} presents the verification and environmental applications, and Sec.~\ref{sec:6} gives the conclusions.

\section{\label{sec:2} Mathematical Formulation}

\begin{figure*}[tbp]
\centering
\includegraphics[width=\textwidth]{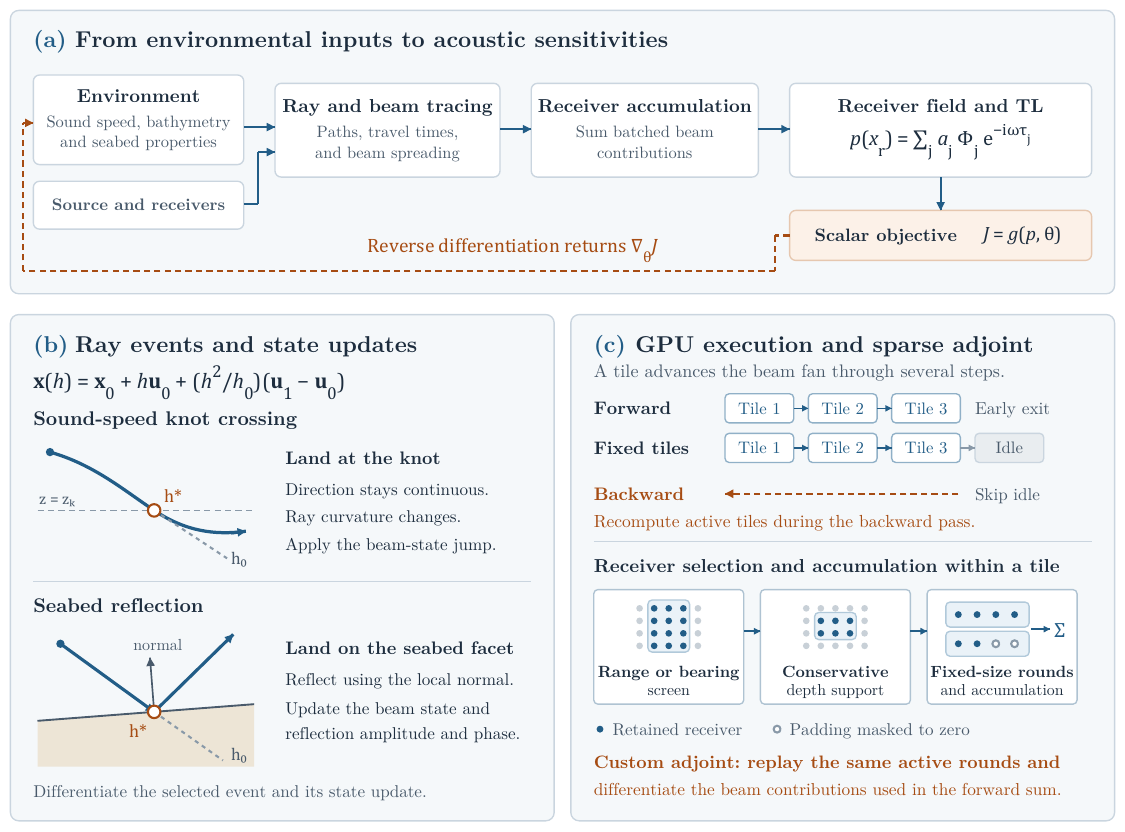}
\caption{DART beam tracing. (a) Acoustic equations and reverse differentiation. (b) Event landing and state updates: accepted (solid) and trial (dashed) paths, contacts (circles). (c) GPU scheduling and sparse adjoint.}
\label{fig:solver}
\end{figure*}

In this section, we present the equations and normalization used in DART, following established ray and beam theory~\cite{jensen2011computational,porter1987gaussian,porter2019beam}. Detailed derivations are given by \citet[Chap.~3]{jensen2011computational} and \citet{cerveny2001seismic}. Figure~\ref{fig:solver} summarizes the acoustic calculation, reverse derivatives, event handling, and GPU execution.

\subsection{Acoustic model and transmission loss}
For a point source at $\mathbf{x}_s$ in a water column of constant density, the frequency-domain pressure $p$ satisfies~\cite[Secs.~2.2 and~2.3.2]{jensen2011computational}
\begin{equation}
\label{eq:helmholtz}
\nabla^2p+\frac{\omega^2}{c^2(\mathbf{x})}p
=-4\pi\delta(\mathbf{x}-\mathbf{x}_s),
\qquad \mathbf{x}\in\Omega,
\end{equation}
where $c$ is the sound speed and $\omega=2\pi f$ is the angular frequency. With the time convention $\operatorname{Re}\{p\,e^{i\omega t}\}$, the outgoing free-field pressure in homogeneous, lossless water is $e^{-i\omega s/c_0}/s$ at distance $s$, where $c_0$ is the homogeneous sound speed. The two-dimensional calculations trace rays in a vertical plane and retain point-source spreading. The seabed reflection coefficient accounts for the density contrast at the interface.

Transmission loss (TL) is referenced to the pressure $p_0$ at 1~m from the same source in homogeneous, lossless water. The point-source strength in Eq.~(\ref{eq:helmholtz}) is set by the factor $4\pi$, which gives a free-field pressure magnitude of $1/s$ for distance $s$ expressed in metres. Thus $|p_0|=1$:
\begin{equation}
\label{eq:TL}
\mathrm{TL}=-20\log_{10}\!\left(\frac{|p|}{|p_0|}\right)
=-10\log_{10}\!\left(\frac{|p|^2}{|p_0|^2}\right).
\end{equation}
Using unit forcing $-\delta$ scales both $p$ and $p_0$ by $1/(4\pi)$ and leaves TL unchanged.

To obtain the ray equations, we apply the high-frequency
Wentzel--Kramers--Brillouin (WKB) approximation away from the source. The leading term for each ray contribution is $A(\mathbf{x})e^{-i\omega\tau(\mathbf{x})}$, where $A$ is the amplitude and $\tau$ is the travel time. Substitution into Eq.~(\ref{eq:helmholtz}) gives the eikonal and transport equations, with ray paths given by the characteristics of the eikonal equation.
We differentiate Eq.~(\ref{eq:TL}) with respect to selected source, frequency, sound-speed, seabed, and bathymetry parameters collected in $\boldsymbol\theta$.

\subsection{Ray paths and beam spreading}
\label{sec:eikonal}
The high-frequency approximation separates travel time $\tau$ from amplitude~\cite[Chap.~3]{jensen2011computational}. With arc length $s$ and slowness $\boldsymbol\sigma=\nabla\tau$, the ray equations in either two or three dimensions are~\cite[Eqs.~(2)--(7) and~(24)--(29)]{porter2019beam}
\begin{equation}
\label{eq:ray}
\frac{d\mathbf{x}}{ds}=c\boldsymbol\sigma,
\qquad \frac{d\boldsymbol\sigma}{ds}=-\frac{\nabla c}{c^2},
\qquad \frac{d\tau}{ds}=\frac{1}{c}.
\end{equation}
Equation~(\ref{eq:ray}) starts at $\mathbf{x}_s$ with $\tau(0)=0$ and $\boldsymbol\sigma(0)=\hat{\mathbf t}_0/c_0$, where $c_0=c(\mathbf{x}_s)$. For declination $\alpha$ and azimuth $\beta$, $\hat{\mathbf t}_0=(\cos\alpha\cos\beta,\cos\alpha\sin\beta,\sin\alpha)$; a vertical-plane calculation uses $(\cos\alpha,\sin\alpha)$.

\label{sec:dynamic}
Dynamic ray tracing supplies beam spreading~\cite{cerveny1982computation,porter1987gaussian}. In a vertical plane, each pair $(q,p)$ obeys~\cite[Eqs.~(10)--(11)]{porter2019beam}
\begin{equation}
\label{eq:paraxial}
\frac{dq}{ds}=cp,
\qquad \frac{dp}{ds}=-\frac{c_{nn}}{c^2}q,
\qquad c_{nn}=\hat{\mathbf n}^{\mathsf T}(\nabla\nabla c)\hat{\mathbf n},
\end{equation}
where $\hat{\mathbf n}$ is the ray normal. We use Porter's normalization: an angular perturbation $\delta\alpha$ gives transverse displacement $\delta n=q_g\delta\alpha/c_0$, with $q_g(0)=0$ and $p_g(0)=1$. In homogeneous water these give $q_g=c_0s$ and $\delta n=s\,\delta\alpha$. The second pair starts at $q_h(0)=1$, $p_h(0)=0$ and is integrated with the same jumps, but the transmission-loss fields reported here use only $q_g$; the second pair enters only an arrival-based field option~\cite{cerveny1982computation,porter2019beam}. All reported fields use geometric hat or Gaussian influences. Sound-speed derivatives in Eq.~(\ref{eq:paraxial}) come from the same interpolant as $c$; gradient discontinuities at profile knots require the explicit jump in Sec.~\ref{sec:diffrules}.

\subsection{Boundary interactions}
For water properties $(\rho_w,c_w)$ and a fluid seabed $(\rho_b,c_b)$, the reflection coefficient at grazing angle $\theta_g$ is~\cite[Chap.~3]{jensen2011computational}
\begin{equation}
\label{eq:rayleigh}
\mathcal R=\frac{\rho_b c_b\csc\theta_t-\rho_w c_w\csc\theta_g}
{\rho_b c_b\csc\theta_t+\rho_w c_w\csc\theta_g},
\qquad \frac{\cos\theta_g}{c_w}=\frac{\cos\theta_t}{c_b}.
\end{equation}
Seabed compressional attenuation $\alpha_p$ enters through a complex sound speed, $\tilde c_b=c_b+i\,a_bc_b^2/\omega$, where $a_b=\alpha_pf/(8.686\,c_b)$ Np m$^{-1}$ for $\alpha_p$ in dB per wavelength (as in all reported cases) and $a_b=\alpha_p/8.686$ for dB m$^{-1}$; the dB-per-wavelength form makes $\operatorname{Im}\tilde c_b$ independent of frequency. In Eq.~(\ref{eq:rayleigh}), $c_b$ is replaced by $\tilde c_b$ and the complex transmission angle is $\sin\theta_t=-i\gamma_b\tilde c_b/\omega$, where $\gamma_b=\sqrt{k_x^2-\omega^2/\tilde c_b^2}$ is the principal root and $k_x=\omega\cos\theta_g/c_w$ is the wavenumber along the boundary; equivalently, $\mathcal R=(\rho_bk_z+i\rho_w\gamma_b)/(\rho_bk_z-i\rho_w\gamma_b)$ with $k_z=\omega\sin\theta_g/c_w$. The transmitted field varies as $e^{-ik_xx-\gamma_b\eta}$ at boundary coordinate $x$ and distance $\eta$ into the seabed. Because $\operatorname{Im}\tilde c_b>0$, both $\operatorname{Re}\gamma_b$ and $\operatorname{Im}\gamma_b$ are positive: the transmitted wave decays into the seabed and carries energy away from the interface, so $|\mathcal R|<1$. Where $\gamma_b$ is imaginary, as for a lossless seabed above the critical angle, the root with $\operatorname{Im}\gamma_b>0$ is taken, as in BELLHOP. In the $e^{-i\omega t}$ convention of \citet{jensen2011computational}, the same seabed has reflection coefficient $\mathcal R^{*}$. The reported cases use Eq.~(\ref{eq:rayleigh}) and a pressure-release surface, $\mathcal R=-1$. Reflection factors accumulate as log-amplitude $\ln\mathcal A$ and phase $\phi$. For boundary normal $\hat{\mathbf n}_b$, specular reflection gives
\begin{equation}
\label{eq:specular}
\hat{\mathbf t}^{+}=\hat{\mathbf t}^{-}
-2(\hat{\mathbf t}^{-}\cdot\hat{\mathbf n}_b)\hat{\mathbf n}_b.
\end{equation}
Equation~(\ref{eq:specular}) updates the direction. In a vertical plane, both dynamic pairs obey the reflection jump~\cite[Sec.~III B, Eqs.~(53)--(54)]{porter2019beam}
\begin{equation}
\label{eq:curvjump}
p_a^{+}=p_a^{-}+\mathcal N_R q_a^{-},
\qquad q_a^{+}=q_a^{-},\qquad a\in\{g,h\}.
\end{equation}
For a planar facet, the coefficient in Eq.~(\ref{eq:curvjump}) is
\begin{equation}
\label{eq:NR}
\mathcal N_R=\frac{2\beta_b^2(\nabla c\cdot\hat{\mathbf n}_b)}{\alpha_b c^2},
\qquad \alpha_b=\hat{\mathbf t}^{-}\cdot\hat{\mathbf n}_b,
\qquad \beta_b=\hat{\mathbf t}^{-}\cdot\hat{\mathbf t}_b,
\end{equation}
where $\hat{\mathbf t}_b$ is the boundary tangent. Equation~(\ref{eq:NR}) is BELLHOP's planar-facet jump after the specular update of Eq.~(\ref{eq:specular}) is substituted; the tangential-gradient terms cancel, so only $\nabla c\cdot\hat{\mathbf n}_b$ remains, even on a sloping facet. It is unchanged if the normal reverses, giving one convention for surface and seabed contacts. The curved-boundary correction, which requires a curvature sign defined relative to $\hat{\mathbf n}_b$, vanishes on planar facets and is used only by the three-dimensional curvilinear option.
\subsection{Receiver influence and field assembly}
\label{sec:influence}

\begin{figure}[t]
  \centering
  \includegraphics[width=\columnwidth]{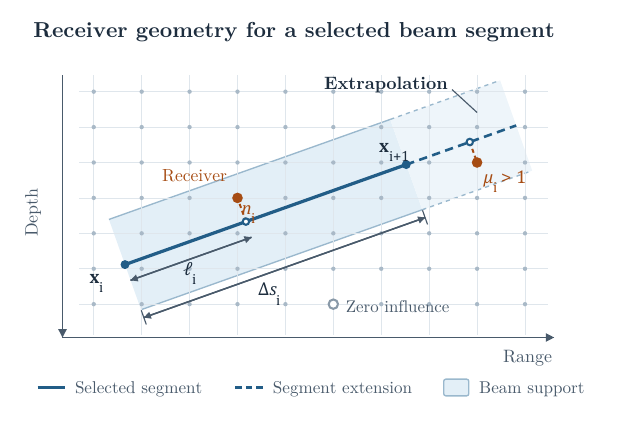}
  \caption{Receiver geometry for a selected beam segment. The along-segment offset $\ell_i$ sets the interpolation coordinate, and the normal offset $n_i$ sets the beam weight. The dashed extension shows extrapolation for $\mu_i>1$.}
  \label{fig:receiver_influence}
\end{figure}

Figure~\ref{fig:receiver_influence} shows how the position of a receiver relative to a selected ray segment determines where we evaluate the beam state and how strongly the beam contributes to the field.

For a segment of length $\Delta s_i$, we use the along-segment offset $\ell_i$ to define $\mu_i=\ell_i/\Delta s_i$. The beam state and travel time are evaluated at the receiver projection:
\begin{equation}
\label{eq:interp}
q_{g,\ast}=(1-\mu_i)q_{g,i}+\mu_iq_{g,i+1},
\qquad
\tau_\ast=(1-\mu_i)\tau_i+\mu_i\tau_{i+1}.
\end{equation}
Other continuous quantities are evaluated in the same way. We allow $\mu_i$ to fall outside $[0,1]$ since clipping it holds the travel time at the endpoint.

The normal offset $n_i$ determines the beam weight. The launch spacing $\delta\alpha$ sets the geometric half-width~\cite[Eq.~(17)]{porter2019beam}
\begin{equation}
\label{eq:width}
W_g=\left|\frac{q_{g,\ast}\delta\alpha}{c_0}\right|,
\qquad
\sigma_W=\max(W_g,W_{\min}).
\end{equation}
The width floor $W_{\min}$ applies only to two-dimensional Gaussian beams near the source and caustics. Section~\ref{sec:diffrules} gives its implementation.

At cylindrical range $\varrho=|R-r_s|$, segment $i$ of beam $j$ contributes~\cite[Eqs.~(6), (18), and~(22)]{porter2019beam}
\begin{equation}
\label{eq:contrib}
\begin{gathered}
p_j(R,z)=\varrho^{-1/2}a_j\Phi(n_i)
e^{i\phi_j}i^{K_j}e^{-i\omega\tau_\ast},\\
a_j=C_\Phi\sqrt{|\cos\alpha_j|}
\sqrt{\frac{c_i}{|q_{g,\ast}|}}\,|\mathcal A_j|,
\end{gathered}
\end{equation}
where $c_i$ is the sound speed at the segment start, $|\mathcal A_j|$ and $\phi_j$ are the accumulated reflection magnitude and phase, and $K_j$ counts sign changes of $q_g$ up to the receiver projection. Hat beams use $C_\Phi=1$ and $\Phi(n)=\max(1-|n|/W_g,0)$. Geometric Gaussian beams use $C_\Phi=(2\pi)^{-1/2}$ and
\begin{equation}
\label{eq:gauss}
\Phi(n)=\frac{W_g}{\sigma_W}
\exp\!\left(-\frac{n^2}{2\sigma_W^2}\right),
\qquad |n|<5\sigma_W,
\end{equation}
with zero influence outside this support. The factor $W_g/\sigma_W$ preserves the integrated transverse weight when the width floor is active. BELLHOP uses a $4\sigma_W$ cutoff, so the cutoffs differ only in the tails of individual beams. Their effect on the relative field error can be larger near coherent nulls. At a caustic node the floored contribution tends to zero as $|q_{g,\ast}|^{1/2}$, but its derivative is singular in the absence of the numerical guard. The derivative checks therefore exclude exact caustic nodes.

We sum complex contributions for the coherent field and accumulate mean-square pressure for the incoherent field:
\begin{equation}
\label{eq:beam_sum}
\begin{aligned}
p &= \sum_{j,i} p_j
&&\text{(coherent)},\\
|p|^2 &= \frac{C_{\mathrm{inc}}}{\varrho}
\sum_{j,i}a_j^2\Phi(n_i)
\left(\frac{W_g}{\sigma_W}\right)^{1/2}
&&\text{(incoherent)}.
\end{aligned}
\end{equation}
The incoherent sum applies $\Phi$ once, following BELLHOP~\cite{acoustics_toolbox}. For two-dimensional Gaussian beams, $C_{\mathrm{inc}}=\sqrt{2\pi}$ and the width ratio differs from unity only where the floor is active. For hat beams, both factors are unity. Squaring the full Gaussian contribution instead produced an incoherent TL bias of about 5.5~dB in our comparison.

When volume absorption is enabled, we apply the BELLHOP factor $\exp(-a c_i\tau)$, with absorption coefficient $a$ in nepers per metre and $c_i\tau$ as an estimate of the propagation distance. The solver supports the Thorp, Francois--Garrison, and Ainslie--McColm models~\cite{thorp1967analytic,francois1982sound,ainslie1998simplified}. Absorption is evaluated separately for each frequency band and disabled in the matched-input comparisons. Section~\ref{sec:beamconv} examines how launch-fan resolution affects the beam sum.

\subsection{Three-dimensional propagation and Nx2D}
\label{sec:3d}

Full three-dimensional propagation advances the ray using Eq.~(\ref{eq:ray}) and evolves beam spreading in two coupled transverse directions. In Porter's transported frame, the dynamic matrices satisfy~\cite[Eqs.~(30)--(36)]{porter2019beam}
\begin{equation}
\label{eq:paraxial3d}
\begin{aligned}[b]
\frac{d\mathbf Q}{ds}&=c\mathbf P,
\qquad
\frac{d\mathbf P}{ds}
=-\frac{\mathbf C_\perp\mathbf Q}{c^2},\\
(\mathbf C_\perp)_{ab}
&=\hat{\mathbf e}_a^{\mathsf T}
(\nabla\nabla c)\hat{\mathbf e}_b,
\end{aligned}
\end{equation}
with $\mathbf Q_g(0)=\mathbf 0$ and $\mathbf P_g(0)=\mathbf I$. The solver uses piecewise-linear, depth-only profiles. Within each layer, $\mathbf C_\perp=\mathbf 0$, and the transverse frame is reconstructed from the ray tangent at zero torsion angle. Profile knots apply the matrix layer jump, and boundary contacts apply coupled jumps in the reflection frame~\cite[Sec.~III B]{porter2019beam}. The reported fields and derivatives use the BELLHOP3D column rotation for $\mathbf P$. On facets tilted out of the ray plane, this convention gives ray-tube widths about 1--2\% different from finite differences. Rotating the components instead agrees with finite differences to $3\times10^{-9}$.

For receiver displacement $\mathbf d$, define $n=|\mathbf d\cdot\hat{\mathbf e}_1|$, $m=|\mathbf d\cdot\hat{\mathbf e}_2|$, and $\tilde{\mathbf Q}=\mathbf Q_g\operatorname{diag}(\delta\alpha,|\cos\alpha|\delta\beta)/c_0$. Following BELLHOP3D, the Gaussian influence is~\cite[Sec.~III C]{porter2019beam,acoustics_toolbox}
\begin{equation}
\label{eq:L1L2}
\begin{aligned}[b]
\begin{pmatrix}a\\b\end{pmatrix}
&=\left|\tilde{\mathbf Q}^{-1}
\begin{pmatrix}n\\m\end{pmatrix}\right|,\\
\Phi_{\mathrm{3D,G}}
&=\exp\!\left(-\frac{a^2+b^2}{2}\right),
\qquad a+b\le4,
\end{aligned}
\end{equation}
with zero influence outside the stated support. The outer absolute value is componentwise. Taking the magnitudes of the offsets before the solve follows the reference and can change the weights when the transverse directions are coupled. The hat weight is $(1-a)(1-b)$ for $\max(a,b)\le1$ and zero elsewhere. Three-dimensional Gaussian beams use no width floor.

The Gaussian amplitude is $A_{\mathrm{3D}}=c|\mathcal A|/(2\pi\sqrt{|\det\mathbf Q_g|})$, and the caustic phase advances by $\pi/2$ at each sign change of $\det\mathbf Q_g$. The incoherent sum is $2\pi\sum A_{\mathrm{3D}}^2\Phi_{\mathrm{3D,G}}$. The discrete Gaussian sum recovers $1/s$ spreading in homogeneous water with a constant 0.20~dB offset, also observed in BELLHOP3D. Launch angles are restricted to $|\alpha|\le89^\circ$.

Nx2D solves independent vertical planes at fixed bearings. It retains in-plane refraction, multipath, and boundary interactions, while neglecting cross-bearing propagation and focusing. The ray paths agree with the full formulation when cross-bearing forcing vanishes. Section~\ref{sec:3dvalid} compares the resulting fields. Both formulations exclude receivers below the local seabed.

\section{\label{sec:3} GPU-resident differentiable solver}

DART implements the acoustic model as one GPU-resident JAX program~\cite{jax2018github}, from beam launch to transmission loss. Explicit differentiation of event locations and interface jumps ensures that sensitivities follow the forward ray and beam dynamics, while sparse accumulation and trajectory recomputation control reverse-mode memory. These choices enable gradients with respect to large environmental parameter sets for the same discrete model used in field verification. Algorithm~\ref{alg:solver} uses two-dimensional notation; the three-dimensional formulation additionally includes azimuthal launch fans and matrix-valued beam influences.

The two-dimensional trajectory uses the state
\begin{equation}
\label{eq:state}
\mathbf Y=[r,z,\xi,\zeta,q_g,q_h,p_g,p_h,\tau,
N_{\mathrm{top}},N_{\mathrm{bot}},\ln\mathcal A,\phi]^{\mathsf T}.
\end{equation}
Here $r$ and $z$ are the horizontal range and depth of the ray, and
$(\xi,\zeta)$ are the corresponding components of the slowness vector
$\boldsymbol{\sigma}$. The pairs $(q_g,p_g)$ and $(q_h,p_h)$ are two
independent solutions of the dynamic ray equations. In each pair, $q$
describes the transverse separation of neighboring rays, and $p$ describes
the corresponding transverse slowness perturbation. The point-source
solution $q_g$ determines the geometric beam width. The variable $\tau$
is the travel time from the source, and $N_{\mathrm{top}}$ and
$N_{\mathrm{bot}}$ count reflections at the sea surface and seabed,
respectively. At each reflection, the logarithm of the reflection
coefficient magnitude is added to $\ln\mathcal A$ and its phase is
added to $\phi$. Together, these variables give the accumulated
complex reflection factor $\mathcal A e^{i\phi}$. Equation~(\ref{eq:state}) has thirteen components, the second dynamic pair being stored as $(q_h,p_h)$. The three-dimensional state has twenty: position (3), slowness (3), the frame torsion angle (1, identically zero for the supported depth-only profiles), the columns of $\mathbf P_g$ (4) and $\mathbf Q_g$ (4), $\tau$, $\ln\mathcal A$, $\phi$, $N_{\mathrm{top}}$, and $N_{\mathrm{bot}}$. Caustic counts $K_j$ are formed from $q_g$ or $\det\mathbf Q_g$ during accumulation and carried across trajectory tiles; they carry no derivative.

\subsection{Differentiation of the discrete acoustic model}
\label{sec:ad}

The parameter vector $\boldsymbol\theta$ contains environmental arrays,
boundary properties, source position, and frequency. We differentiate the
same propagation steps, interface jumps, interpolation, and beam influence
calculations used to assemble the forward field. The launch map $\Psi$
initializes the state, and the step map $\mathcal{S}$ advances it. For
$N_s$ propagation steps, the forward calculation is
\begin{equation}
\label{eq:recurrence}
\begin{aligned}[b]
\mathbf Y^0&=\Psi(\boldsymbol\theta),\\
\mathbf Y^k&=\mathcal{S}(\mathbf Y^{k-1},\boldsymbol\theta),\\
\mathcal J&=g(\mathbf Y^0,\ldots,\mathbf Y^{N_s},\boldsymbol\theta).
\end{aligned}
\end{equation}
The scalar objective $\mathcal J$ can depend on the full trajectory. For
example, consider an objective based on the acoustic field in a receiver
window. A ray state can affect the objective directly through its
contribution to that field and indirectly through the later states it
produces. The adjoint $\boldsymbol\lambda^k$ combines both effects as
the chain rule is applied backward~\cite{griewank2008evaluating,plessix2006review}:
\begin{equation}
\label{eq:adjoint_recur}
\boldsymbol\lambda^{k-1}
=\left(\frac{\partial\mathcal{S}}{\partial\mathbf Y^{k-1}}\right)^{\!\mathsf T}
\boldsymbol\lambda^k
+\left(\frac{\partial g}{\partial\mathbf Y^{k-1}}\right)^{\!\mathsf T},
\qquad k=N_s,\ldots,1,
\end{equation}
with terminal value
$\boldsymbol\lambda^{N_s}=(\partial g/\partial\mathbf Y^{N_s})^{\mathsf T}$.
The step derivatives are evaluated along the forward trajectory.

The parameter gradient combines the direct dependence of the objective
on $\boldsymbol\theta$ with its effects on the propagation steps and the
launch state:
\begin{equation}
\label{eq:param_grad}
\nabla_{\boldsymbol\theta}\mathcal J
=\left(\frac{\partial g}{\partial\boldsymbol\theta}\right)^{\!\mathsf T}
+\sum_{k=1}^{N_s}
\left(\frac{\partial\mathcal{S}}{\partial\boldsymbol\theta}\right)^{\!\mathsf T}
\boldsymbol\lambda^k
+\left(\frac{\partial\Psi}{\partial\boldsymbol\theta}\right)^{\!\mathsf T}
\boldsymbol\lambda^0.
\end{equation}
The partial derivative of $g$ with respect to $\boldsymbol\theta$ holds
the trajectory states fixed. The remaining terms account for how those
states change with the parameters.

Reverse differentiation evaluates the transposed Jacobian products in
Eqs.~(\ref{eq:adjoint_recur}) and~(\ref{eq:param_grad}) without forming the
matrices. A single reverse pass gives the gradient of one scalar objective
with respect to all input parameters, with arithmetic cost that scales
with the forward calculation~\cite{griewank2008evaluating}. Checkpointing
reduces memory use by retaining selected states and recomputing the
intervening steps during the backward pass. Forward differentiation
propagates one chosen parameter perturbation and can return its effect at
every receiver. A full field Jacobian generally requires multiple forward
directions or reverse passes.

The derivatives describe local changes for a fixed sequence of events
and fixed layer, facet, and receiver-support selections. Event locations
can move and are differentiated within that sequence. When a discrete
selection changes, the gradient may jump or be undefined at the
transition. The finite-perturbation tests in Sec.~\ref{sec:kernels} assess
how far these local derivatives predict the effects of environmental
changes.

\subsection{Event locations, interface jumps, and numerical decisions}
\label{sec:diffrules}
Event handling determines the state passed to every subsequent step (Fig.~\ref{fig:solver}b). We therefore locate events on the numerical trajectory and differentiate their locations together with the state update. For a trial step $h_0$, the modified-polygon Runge--Kutta scheme in BELLHOP~\cite{porter2019beam,acoustics_toolbox} gives
\begin{equation}
\label{eq:rk2}
\mathbf Y(h)=\mathbf Y_0+h[(1-w)\mathbf f_0+w\mathbf f_1],
\qquad w=h/h_0,
\end{equation}
where $\mathbf f_0=\mathbf f(\mathbf Y_0)$ and $\mathbf f_1=\mathbf f(\mathbf Y_{1/2})$, with $\mathbf Y_{1/2}=\mathbf Y_0+\tfrac12h_0\mathbf f_0$. Equation~(\ref{eq:rk2}) is second-order accurate for every $h\le h_0$ and reduces to the midpoint rule at $h=h_0$. With $\mathbf u_0$ and $\mathbf u_1$ the position rates $c\boldsymbol\sigma$ of $\mathbf f_0$ and $\mathbf f_1$, the position follows the quadratic path
\begin{equation}
\label{eq:quadratic}
\mathbf x(h)=\mathbf x_0+h\mathbf u_0+\frac{h^2}{h_0}(\mathbf u_1-\mathbf u_0).
\end{equation}
For a planar event surface, the signed distance along Eq.~(\ref{eq:quadratic}) is $f_0+\beta h+\gamma h^2$, and we take the smallest admissible root with outward motion, $f'(h^\ast)>0$, from the cancellation-free pair $f_0/\varkappa$ and $\varkappa/\gamma$, where $\varkappa=-\tfrac12[\beta+\operatorname{sgn}(\beta)\sqrt{\beta^2-4\gamma f_0}]$. The root $f_0/\varkappa$ approaches $-f_0/\beta$ as $\gamma\to0$, allowing the linear limit to be evaluated without dividing by $\gamma$. Negative discriminants are rejected to avoid false reflections at turning points. Without an event, $h^\ast=h_0$, and the selected root remains in the differentiated calculation. In three dimensions, profile knots, seabed facets, cell edges, and cell diagonals are all landed in this way. In two dimensions, top and bottom contacts are landed in this way, whereas profile knots and bathymetry breakpoints use BELLHOP's straight-line step reduction, applied before and after the half step~\cite{acoustics_toolbox}; a residue within 32 units in the last place is then snapped onto the knot. Snapping replaces the coordinate by the interface value, so the normal coordinate inherits that interface's derivative, which is zero for a fixed knot.

In two dimensions, reflections apply Eq.~(\ref{eq:NR}) together with the specular direction update. At a horizontal sound-speed gradient discontinuity, each dynamic pair receives the weak-interface jump~\cite[Sec.~III B]{porter2019beam}, implemented as in BELLHOP~\cite{acoustics_toolbox}
\begin{equation}
\label{eq:layerjump}
p^+=p-q\,\frac{R_M(2c_{n,\mathrm{jump}}-R_Mc_{s,\mathrm{jump}})}{c},
\qquad R_M=\frac{t_r}{t_z}.
\end{equation}
In Eq.~(\ref{eq:layerjump}), $(t_r,t_z)=(\xi,\zeta)$ are slowness components and $\Delta\nabla c$ is the gradient change from incident to crossed layer. The projections $c_{n,\mathrm{jump}}=\Delta\nabla c\cdot(-t_z,t_r)$ and $c_{s,\mathrm{jump}}=\Delta\nabla c\cdot(t_r,t_z)$ therefore use slowness vectors rather than unit vectors and have units of m$^{-1}$. For a horizontal knot, $\Delta\nabla c=(0,\Delta c_z)$, and Eq.~(\ref{eq:layerjump}) reduces to $p^+-p=-q\,\Delta c_z\cos^2\theta/(c^2\sin\theta)$, where $\theta$ is the ray angle at the knot; the coefficient therefore diverges as $1/\sin\theta$ for rays tangent to the knot. Reflection and layer tests match finite-difference ray-tube widths to about $10^{-9}$ and $10^{-8}$, respectively. A two-dimensional source exactly on a knot is assigned to the layer above, its first step keeps that layer for both stages, is cut to $10^{-3}\,ds$ if it leaves through the knot, and then receives the jump of Eq.~(\ref{eq:layerjump}). Likewise, a trajectory segment outside the receiver range grid contributes to no receiver column. 

Rounding must not change the event sequence under small input perturbations. We clamp the three-dimensional knot search to interior knots so that a surface contact displaced by rounding cannot carry the wrong layer gradient into the next step. In a shallow-source test the perturbation-induced and CPU--GPU field differences then fall from 0.7 dB median to about $10^{-4}$ dB. Masks and structured array primitives keep data-dependent control flow inside the compiled program; discrete indices and bounce counts carry no derivative. 

Reproducing the reference Gaussian fields at matched inputs also requires using the same minimum beam width in Eq.~(\ref{eq:width}). This width floor limits how narrow a beam can become in the field calculation and therefore affects its contribution at nearby receivers. Matching the ray trajectories alone is consequently insufficient to reproduce the reference field; the beam-width convention must also be retained.

BELLHOP specifies this floor through an empirical rule applied to numerical values expressed in SI units. To state that convention explicitly, we define $f_{\mathrm{Hz}}=f/(1\,\mathrm{Hz})$, $\tau_{\mathrm s}=\tau/(1\,\mathrm s)$, and $\lambda_{\mathrm m}=\lambda/(1\,\mathrm m)$, where $\tau$ is the end-segment travel time and $\lambda=c_0/f$ is the wavelength evaluated using $c_0$; the reference evaluates $\lambda$ with the sound speed at the segment start. The minimum width is then
\begin{equation}
\label{eq:widthfloor}
W_{\min}
=(1\,\mathrm m)\min\left(
\frac{0.2f_{\mathrm{Hz}}\tau_{\mathrm s}}{\lambda_{\mathrm m}},
\pi\lambda_{\mathrm m}
\right).
\end{equation}
The first branch increases with end-segment travel time at fixed frequency and wavelength, while the second caps the floor at $\pi\lambda$. The normalized quantities inside the minimum are dimensionless numerical values, and the prefactor restores the width in metres. This notation records the BELLHOP implementation's empirical numerical convention explicitly; it should not be interpreted as a physical scaling law.

Retaining the precise rule in Eq.~(\ref{eq:widthfloor}) matters for field agreement. An implementation that used $0.2f\tau$ in the floor left two-dimensional Gaussian beams on the 50 Hz Munk slice 2.6 dB RMS from BELLHOP, against 0.11 dB with Eq.~(\ref{eq:widthfloor}), showing that the width floor can materially affect the predicted field even when the environmental inputs are unchanged. Section~\ref{sec:beamconv} further examines sensitivity to the launch fan, assessing how the field responds to changes in angular sampling with the reference width convention retained.

\subsection{Sparse accumulation and its custom adjoint}
\label{sec:sparse}
\begin{figure*}[tbp]
\centering
\includegraphics[width=\textwidth]{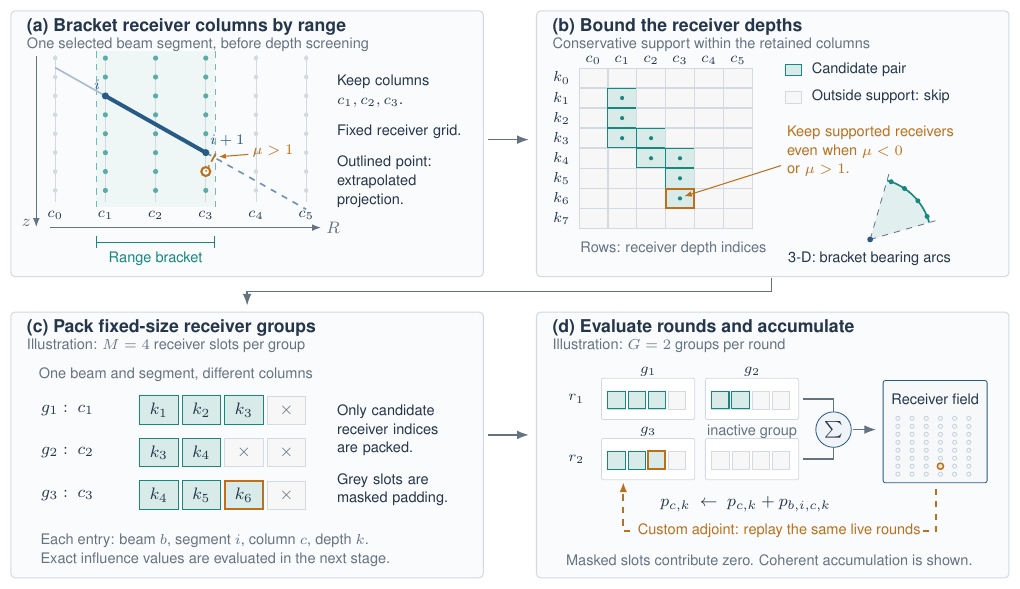}
\caption{Sparse beam--receiver accumulation. (a) Range brackets select receiver columns. (b) Conservative depth supports retain candidates, including extrapolated projections. (c) Receiver indices are packed into fixed-size groups with masked padding. (d) Fixed-capacity rounds accumulate the coherent field; the custom adjoint replays the selected rounds.}
\label{fig:sparse_accumulation}
\end{figure*}

Most beam--receiver pairs make no contribution to the field. Evaluating
all pairs adds work and creates intermediate values that are not needed in the reverse pass. The sparse scheduler in Fig.~\ref{fig:solver}c first identifies receiver columns by range in two dimensions and by bearing arcs in three dimensions. It then computes conservative depth bounds that account for the extrapolation in Eq.~(\ref{eq:interp}). Within these bounds, receivers are grouped by depth and evaluated in fixed-size rounds. Figure~\ref{fig:sparse_accumulation} shows how receivers are selected, grouped, and evaluated.
Each trajectory tile contains 64 steps, and each receiver group contains
$\min(32,N_z)$ depths. The round capacity is fixed at a power of two based on the expected entry count. It is 32,768 entries for Dickins and 65,536 for Munk. The scheduler repeats rounds until all supported entries have been evaluated, so the capacity limits the work per round without truncating the field. The number of receiver columns allocated per segment is also fixed at $\lfloor h_{\max}/\Delta r\rfloor+1$ in two dimensions and $\lfloor h_{\max}/\Delta r\rfloor+2$ in three dimensions. These bounds are based on the maximum step length. The column bounds and trajectory step budget are set before execution and are not checked at run time.

The custom accumulation adjoint replays the live rounds selected in the forward pass and differentiates the same influence expression, Eq.~(\ref{eq:contrib}). The integer schedule of a tile is kept as a non-differentiable residual. Because accumulation is a scatter-add, each round's backward pass gathers the field cotangent at that round's receivers and pulls it back to the contributing segments through the influence expression. This keeps the pressure and gradient calculations consistent while limiting working trajectory and influence intermediates to one step tile and one receiver round; parameter cotangents, including per-beam boundary cotangents, are not bounded by this schedule. Table~\ref{tab:implspeed} summarizes the measured reductions in work and runtime.

We compare sparse accumulation with an uncut evaluation of the same beam-influence formula. The reference evaluates every traced segment against all receiver depths on its bracketed range columns, without depth culling or receiver rounds. CPU tests use 999 hat beams for Munk and 181 Gaussian beams for Dickins. The maximum TL differences are $1.3\times10^{-10}$ and $9.8\times10^{-10}$~dB, respectively, showing agreement to numerical precision in these configurations.

For large three-dimensional kernels, boundary cotangents still scale with the product of beam and bathymetry-node counts. We split the fan into $B$ azimuthal blocks, $p=\sum_b p_b$, and hold the full coherent field $\bar p=p$ fixed during each block's differentiation. For a receiver window $W$,
\begin{equation}
\label{eq:blocks}
\frac{\partial\,\mathrm{TL}_W}{\partial d_{kl}}
=-\frac{20}{\ln 10}\,\frac{1}{N_W\,\langle|p|^{2}\rangle_W}
\sum_{b=1}^{B}\frac{\partial}{\partial d_{kl}}\sum_{i\in W}\operatorname{Re}\!\left(\bar p_i^{*}\,p_{i,b}\right),
\end{equation}
where $\mathrm{TL}_W=-10\log_{10}(\langle|p|^2\rangle_W/|p_0|^2)$, $d_{kl}$ is node depth, and $N_W$ is the receiver count. Equation~(\ref{eq:blocks}) preserves interference between blocks through the full field $\bar p$, while only one block's cotangents are evaluated at a time. Summed block fields agree with the full fan to better than $10^{-12}$.

\subsection{GPU execution and trajectory reuse}
\label{sec:impl}
We trace independent beams in batches on the GPU~\cite{kirk2016programming}. Trajectory arrays are stored in (steps, beams, state) order so that receiver accumulation can use them directly. Avoiding a full transpose gave a $2.3\times$ speedup in an isolated comparison of the two layouts. The compiler handles thread assignment and memory coalescing, and we did not profile either directly. Each Runge--Kutta stage uses a sorted search to look up tabulated environmental values. The default JAX search runs a loop of about $\log_2 N$ iterations for a table with $N$ entries. On the GPU, we use a loop-free version of the same search that returns identical indices.
In the compiled Dickins calculation, this change reduced the loop count
from 93 to 4. Of the original 93 loops, 89 were inside the tracing step.
These counts describe the compiled program. We did not measure kernel
launches directly. On the CPU, we retain binary search.
We use CUDA graphs to reduce launch overhead and unroll the tracing loop
so that each compiled iteration advances four steps. A persistent cache
reuses compiled programs across runs and reduces cold-start time.
Table~\ref{tab:implspeed} summarizes the measured performance.

\begin{table*}[!htbp]
\caption{Measured performance on the hardware described in
Sec.~\ref{sec:setup}. Dense accumulation evaluates all beam--receiver
pairs.}
\label{tab:implspeed}
\begin{ruledtabular}
\begin{tabular}{@{}p{0.20\textwidth}p{0.29\textwidth}p{0.44\textwidth}@{}}
Component & Implementation & Measured result \\
\hline
Receiver accumulation
  & Sparse scheduler, Sec.~\ref{sec:sparse}
  & Evaluates 1--4\% of the dense beam--receiver pairs.
    Maximum TL differences from an uncut reference are below
    $10^{-9}$~dB for Munk (999 hat beams) and Dickins
    (181 Gaussian beams). \\
Tabulated lookups
  & Loop-free sorted search
  & $10.9\times$ speedup for GPU tracing relative to the same
    program with iterative binary search. \\
GPU execution
  & Full solver on the GPU
  & $1.0$--$10.8\times$ speedup across tested forward schedules
    and gradients relative to the same JAX program on 16 CPU cores. \\
Compilation
  & Persistent compilation cache
  & Cold-start Munk solve reduced from 15.5 to 11.5~s. \\
Broadband prediction
  & One trajectory trace shared across frequencies
  & Ten frequencies evaluated at the cost of 2.0--2.5
    single-frequency forward solves. \\
Three-dimensional kernels
  & Azimuthal blocking, Eq.~(\ref{eq:blocks})
  & A 40,000-node kernel computed on a GPU with 8~GB of memory. \\
\end{tabular}
\end{ruledtabular}
\end{table*}

The execution schedule also controls the memory required for reverse differentiation. A trajectory tile contains a fixed number of consecutive integration steps across the beam fan. Forward evaluation exits when all beams are inactive. Reverse mode uses a fixed sequence of tiles, skips idle tiles, and recomputes active tiles during the backward pass~\cite{griewank2008evaluating}. Each recomputed tile rebuilds its receiver schedule from the recomputed trajectory, and its accumulation is differentiated by the custom adjoint of Sec.~\ref{sec:sparse}, so schedules are not stored across tiles. Recomputation avoids retaining every intermediate operation for the full trajectory, trading additional arithmetic for lower memory use. The forward and reverse-mode schedules agree to about $10^{-14}$ in the field.

When the medium and launch fan are independent of frequency, we trace
the rays once and reuse the segments within each trajectory tile. The
beam width, phase, and receiver influence are evaluated separately at
each frequency. This avoids repeating the ray integration or storing
a full trajectory for each frequency. The shared-trace mode currently
supports forward evaluation only.
Reflection factors are evaluated at the band-centre frequency and reused
across the band. This reuse is exact when seabed attenuation expressed
per wavelength is frequency independent and is approximate when the
reflection factors vary with frequency. Tracing the rays separately at
$f_0$, $2f_0$, and $f_0/2$ gave identical trajectory states for Dickins
and changes of about $10^{-11}$ in the Munk reflection amplitude and
phase. The frequency sweep below assesses trajectory reuse under
these conditions.

\begin{algorithm}[tbp]
\currcaption{\label{alg:solver}DART transmission-loss solve, shown for a coherent two-dimensional field; incoherent runs accumulate Eq.~(\ref{eq:beam_sum}) instead. Here $k$ indexes beams and $n$ steps within a tile.}
\KwIn{Source $(r_s,z_s)$, frequency $f$, launch angles $\{\alpha_k\}_{k=1}^{N_\alpha}$; environment $\boldsymbol{\theta}=(c(\mathbf{x}),$ bathymetry, seabed$)$; receiver grid $(R,z)$; step $ds$, steps $N_s$}
\KwOut{Transmission-loss field $\mathrm{TL}(R,z)$; gradient $\nabla_{\boldsymbol{\theta}}\mathcal{J}$ of a scalar objective $\mathcal{J}$}
\BlankLine
Initialize beam-fan states $\mathbf{Y}^0_k$ for all $\alpha_k$ \tcp*[r]{parallel over beams (SIMT)}
\For{each tile of $T$ steps}{
  \For{$n$ in the tile}{
    $h_0 \leftarrow$ trial step; $h^\ast \leftarrow$ first admissible event step from Eq.~(\ref{eq:quadratic}), or $h_0$\;
    $\mathbf{Y}^n \leftarrow \mathrm{RK2step}(\mathbf{Y}^{n-1}, h^{\ast})$ \tcp*[r]{lands on the event}
    apply the layer jump~(\ref{eq:layerjump}), or the reflection Eqs.~(\ref{eq:rayleigh}), (\ref{eq:specular}), and~(\ref{eq:NR}), by mask \tcp*[r]{branchless}
  }
  bracket (beam, segment, receiver-column) entries; build conservative supports\;
  evaluate Eq.~(\ref{eq:contrib}) in fixed-size rounds; accumulate into $p(R,z)$ \tcp*[r]{sparse}
}
$\mathrm{TL}(R,z) \leftarrow -20\log_{10}\!\big(|p|/|p_0|\big)$; mask receivers below the seafloor\;
\BlankLine
$\mathcal{J} \leftarrow$ scalar reduction of $\mathrm{TL}$ (e.g.\ a window-level loss)\;
$\nabla_{\boldsymbol{\theta}}\mathcal{J} \leftarrow$ reverse pass: tiles recomputed, rounds replayed by the custom adjoint \tcp*[r]{one pass}
\end{algorithm}

\section{\label{sec:5} Results and Discussion}

We compare acoustic fields and derivatives with BELLHOP calculations, then examine environmental sensitivity and synthetic inversion.

\subsection{Numerical setup and reproducibility}
\label{sec:setup}
\begin{figure*}[tbp]
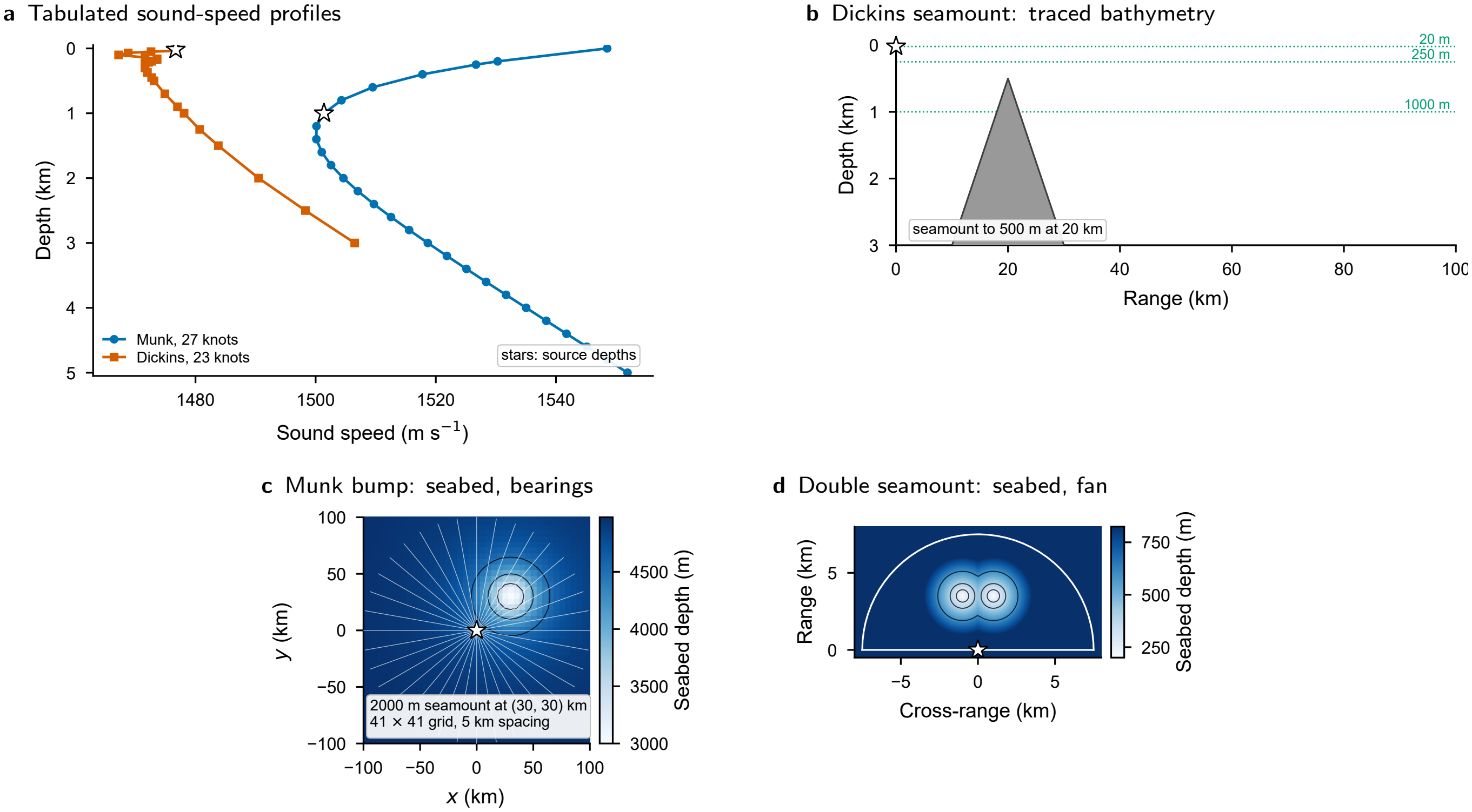

\centering
\pnl{81mm}{a}{Tabulated sound-speed profiles}{panels/env_ssp.pdf}\hfill
\pnl{81mm}{b}{Dickins seamount: traced bathymetry}{panels/env_dickins.pdf}\\[2mm]
\pnl{53.5mm}{c}{Munk bump: seabed, bearings}{panels/env_munkbump.pdf}\hspace{8mm}
\pnl{53.5mm}{d}{Double seamount: seabed, fan}{panels/env_dseam_clean.pdf}
\caption{Environments of the reference cases. (a)
Sound-speed profiles used by both codes. (b) The Dickins bathymetry, rising to 500 m at 20 km, with the dotted receiver depths. (c) The Munk bump, a 2000 m Gaussian seamount at $(30, 30)$~km. (d) The double seamount in the frame of Fig.~\ref{fig:nx2d}, with the $\pm90^\circ$ receiver fan to 7.5 km.}
\label{fig:cases}
\end{figure*}

In this section, we describe the benchmark environments and the
discretizations used in the numerical comparisons.
Figure~\ref{fig:cases} shows the two tabulated sound-speed profiles,
the Dickins bathymetry, and the two three-dimensional seabeds with
their receiver fans. Table~\ref{tab:setup} lists the corresponding
discretizations. We use the same launch fan, angular aperture, and step
size in each comparison. The Nx2D tests also use the same seabed facets.
Matching these inputs allows us to compare the implementations at the
same discretization.
All cases represent the seabed as a fluid half-space, with attenuation
specified in dB per wavelength. The Munk cases use a seabed sound speed
of 1600~m~s$^{-1}$, a density of 1.8~g~cm$^{-3}$, and an attenuation
of 0.8~dB/$\lambda$. The corresponding values for Dickins are
1550~m~s$^{-1}$, 1.5~g~cm$^{-3}$, and 0.5~dB/$\lambda$.
For the double-seamount case, the water has a constant sound speed of
1500~m~s$^{-1}$, and the seabed parameters are 1650~m~s$^{-1}$,
1.5~g~cm$^{-3}$, and 0.5~dB/$\lambda$.

Receiver grids span 0--100 km in range with 1001 ranges for the two-dimensional cases, with 501 depths to 5000 m for Munk and 601 depths to 3000 m for Dickins. The Munk slice uses the same 501 depths and 1001 ranges on two bearings. The Munk bump uses 1001 ranges at 1000 m depth on 36 bearings. The double seamount uses 1501 ranges to 7.5 km at 400 m depth on 721 bearings from $-90^\circ$ to $90^\circ$. The Munk-bump and double-seamount bathymetries are the reference test grids: a $41\times41$ grid of 5 km cells and a $1000\times1000$ grid of 16 m cells, respectively, of which Fig.~\ref{fig:cases}d shows the receiver-fan subdomain.

For a full-circle grid, $0^\circ$ and $360^\circ$ represent the same direction. We therefore omit the final endpoint to avoid counting that direction twice. In the Munk-bump case, the input requests 144 bearings from $0^\circ$ to $360^\circ$. Removing the duplicate leaves 143 unique azimuths, each with 200 declination angles, giving $200\times143=28{,}600$ beams. The receiver grid follows the same convention: 37 requested bearings become 36 unique azimuths, giving 36,036 receivers in total.
The Munk-slice receivers are placed at bearings of $0^\circ$ and $1^\circ$. In the Nx2D calculations, beams are traced within separate vertical planes aligned with the receiver bearings, so no azimuth fan is used.

The reference codes are BELLHOP and BELLHOP3D from the Acoustics Toolbox release of 29 August 2026~\cite{acoustics_toolbox}, compiled with GNU Fortran 13.3.0 using \texttt{-O3 -march=native}, \texttt{-ffast-math}, \texttt{-funroll-all-loops}, \texttt{-fomit-frame-pointer}, \texttt{-std=gnu} and \texttt{-w}, and run on one CPU core. Both receive the same tabulated Munk profile and knot interfaces. The solver uses double precision on an 8 GB NVIDIA RTX 2000 Ada laptop GPU (driver 573.24) under WSL2, JAX 0.11.2, and CUDA 12.8. Its double-precision rate is one sixty-fourth of its single-precision rate. Inversion and reference runs are performed with an Intel Core Ultra 7 165H. Forward times compared with the reference codes (Sec.~\ref{sec:speed}) are medians of 3--7 warm runs of each code, excluding the first run; the two-dimensional gradient and frequency-sweep costs are medians of two warm runs, and the bathymetry-gradient scaling uses medians of three. Compilation is excluded and reported separately.

\begin{table*}[tbp]
\caption{Case discretizations: frequency, source depth, declination $\alpha$, azimuth $\beta$, traced fan, and nominal step. Three-dimensional fans list traced declination $\times$ azimuth counts, after removing the duplicate $360^\circ$ endpoint of full-circle fans. Matched comparisons use the same seabed facets; receiver counts appear in Tables~\ref{tab:accuracy} and~\ref{tab:nx2d}.}
\label{tab:setup}
\setlength{\tabcolsep}{4pt}
\begin{ruledtabular}
\begin{tabular}{llrrrrrr}
Case & Mode & $f$ (Hz) & $z_s$ (m) & $\alpha$ (deg) & $\beta$ (deg) & fan & $ds$ (m) \\
\hline
Munk & 2-D, hat & 50 & 1000 & $\pm20.3$ & \textendash & 999 & 50 \\
Dickins & 2-D, Gaussian & 230 & 18 & $\pm45$ & \textendash & 4672 & 5 \\
\hline
Munk slice & 3-D, Gaussian & 50 & 1000 & $\pm20.3$ & $\pm2$ & $501\times51$ & 100 \\
Munk bump & 3-D, hat & 50 & 1000 & $-14.66$ to $20$ & $0$--$360$ & $200\times143$ & 100 \\
Double seamount & 3-D, Gaussian & 100 & 250 & $\pm80$ & $\pm90$ & $100\times113$ & 20 \\
\hline
Munk slice & Nx2D, Gaussian & 50 & 1000 & $\pm20.3$ & \textendash & 501 & 100 \\
Munk bump & Nx2D, hat & 50 & 1000 & $-14.66$ to $20$ & \textendash & 200 & 100 \\
Double seamount & Nx2D, Gaussian & 100 & 250 & $\pm80$ & \textendash & 100 & 20 \\
\hline
Inversion data & 3-D, incoherent & 100 & 250 & $\pm80$ & $\pm90$ & $200\times226$ & 10 \\
\end{tabular}
\end{ruledtabular}
\end{table*}

\subsection{Agreement with BELLHOP in two dimensions}
\label{sec:parity2d}
\begin{figure*}[tbp]
\centering
\includegraphics{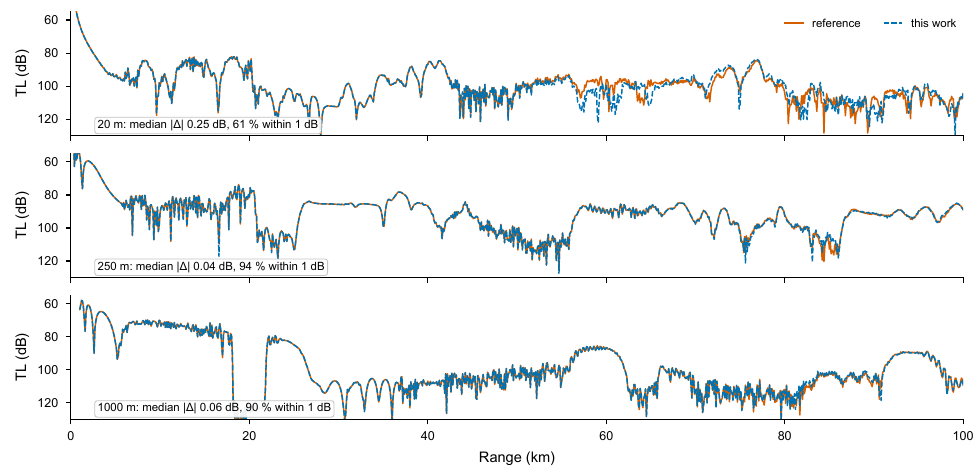}
\caption{Dickins range profiles at 20, 250, and 1000 m: DART (dashed) and BELLHOP (solid).}
\label{fig:slices}
\end{figure*}

With 999 Munk and 4672 Dickins beams at matched inputs, RMS differences from BELLHOP are 0.51 and 1.16 dB for receivers with TL below 150 dB in both fields (Fig.~\ref{fig:fidelity}a--d; Table~\ref{tab:accuracy}). Median absolute differences are 0.023 and 0.049 dB, with 99.4 and 92.5 percent within 1 dB. The larger RMS values reflect localized residuals, including Dickins surface arrivals beyond 50 km and interference nulls, where pressure cancellation makes TL sensitive to phase. At 20, 250, and 1000 m depth, median differences are 0.25, 0.04, and 0.06 dB; 61, 94, and 90 percent of range samples lie within 1 dB (Fig.~\ref{fig:slices}).

\begin{figure*}[tbp]
\centering
\pnl{81mm}{a}{Munk, 2-D: TL (999 hat beams, 50 Hz)}{panels/fid_munk_tl.pdf}\hfill
\pnl{81mm}{b}{Munk: difference from the reference}{panels/fid_munk_diff.pdf}\\[2mm]
\pnl{81mm}{c}{Dickins, 2-D: TL (4672 Gaussian beams, 230 Hz)}{panels/fid_dickins_tl.pdf}\hfill
\pnl{81mm}{d}{Dickins: difference from the reference}{panels/fid_dickins_diff.pdf}\\[2mm]
\pnl{53.5mm}{e}{Parity at every fan size}{panels/fid_sweep.pdf}\hfill
\pnl{53.5mm}{f}{Differences, all five cases}{panels/fid_hist.pdf}\hfill
\pnl{53.5mm}{g}{Run time}{panels/fid_time.pdf}
\caption{Matched-input verification of DART. (a), (b) Munk TL and difference from BELLHOP. (c), (d) Dickins fields; grey denotes seabed. (e) Inter-code RMS differences at equal fan sizes (solid) and reference changes relative to its finest fan (dashed). (f) Residual distributions; legend percentages are the fractions of illuminated receivers within 1 dB. (g) Warm-run times of the parity runs.}
\label{fig:fidelity}
\end{figure*}

\subsection{Agreement across launch fans}
\label{sec:beamconv}

In this subsection, we examine whether the two solvers agree as the
number of launch beams changes. For the Munk case, we test fans with
61--2001 beams. The BELLHOP field differs from its finest-fan result
by up to 4.2~dB RMS, while the RMS difference between the solvers at
matched beam counts remains between 0.44 and 0.52~dB
(Fig.~\ref{fig:fidelity}e). The Dickins case shows the same pattern
over 181--4672 beams. The BELLHOP field differs from its finest-fan
result by up to 8.8~dB RMS, while the differences between the solvers
range from 0.81 to 1.80~dB.

Increasing the beam count reduces the spacing between launch angles
and changes the beam width through Eq.~(\ref{eq:width}). The two
solvers can therefore agree at each beam count even while both fields
continue to change as more beams are added. Convergence would require
the field to approach a stable result as the fan is refined. Here,
the finest fan is simply the one with the most beams tested: 2001
for Munk and 4672 for Dickins. Its field provides a reference for
measuring changes within this comparison.


\subsection{Three-dimensional propagation and the Nx2D approximation}
\label{sec:3dvalid}

In this subsection, we compare full three-dimensional predictions with
BELLHOP3D and examine how the Nx2D approximation affects the acoustic
field. For each environment, we first compare the full three-dimensional
solvers at matched inputs. We then compare the Nx2D implementations and
examine the effects of seabed discretization and reflection. Finally,
we compare the reference Nx2D and full three-dimensional fields.
Tables~\ref{tab:accuracy} and~\ref{tab:nx2d} summarize these comparisons.
Across the three matched full three-dimensional cases, at least
99.6\% of illuminated receivers agree within 1~dB. The largest grid
contains about 1.1 million receivers.

\begin{figure*}[tbp]
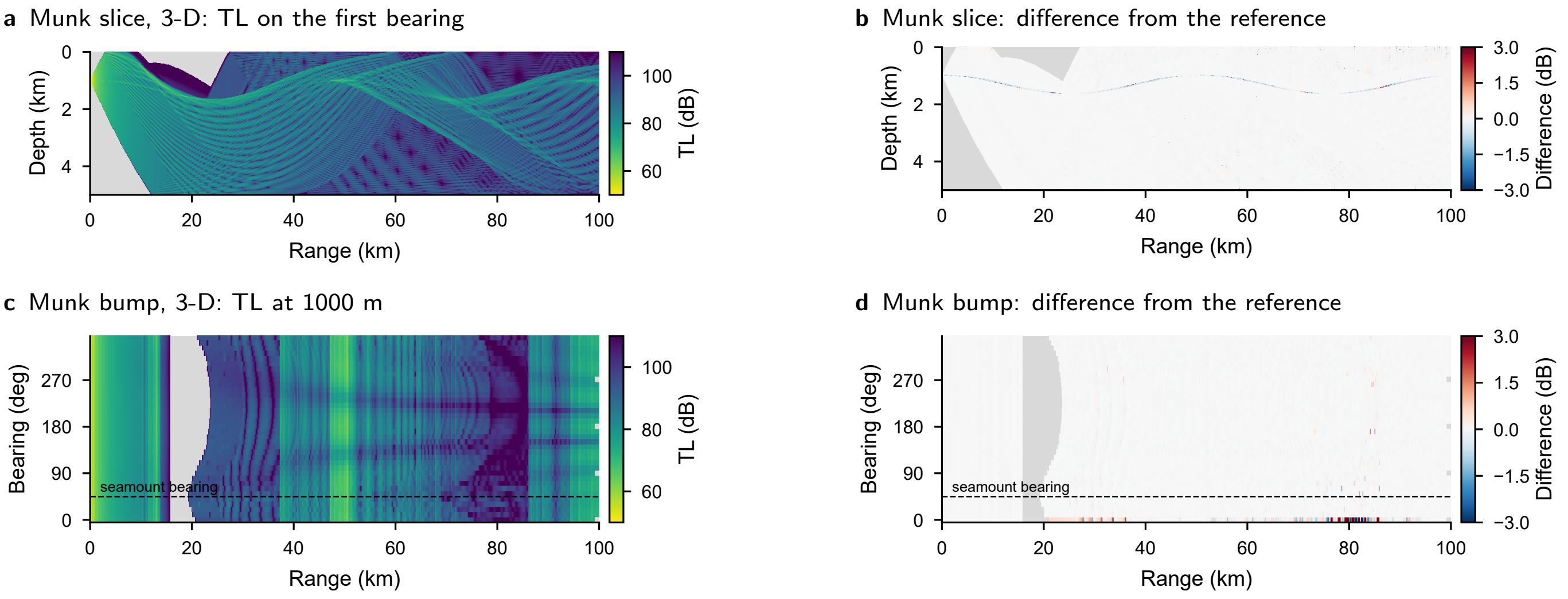

\centering
\pnl{81mm}{a}{Munk slice, 3-D: TL on the first bearing}{panels/fid3d_munkslice_tl.pdf}\hfill
\pnl{81mm}{b}{Munk slice: difference from the reference}{panels/fid3d_munkslice_diff.pdf}\\[2mm]
\pnl{81mm}{c}{Munk bump, 3-D: TL at 1000 m}{panels/fid3d_munkbump_tl.pdf}\hfill
\pnl{81mm}{d}{Munk bump: difference from the reference}{panels/fid3d_munkbump_diff.pdf}
\caption{Matched 3-D TL and differences from BELLHOP3D.
(a,b) Munk slice on the first bearing. (c,d) Munk bump at 1000~m
depth across 36 bearings. The dashed line marks the seamount bearing.
Grey indicates receivers below the coverage floor.}
\label{fig:fid3d}
\end{figure*}

The Munk slice provides a baseline with a flat seabed at 5000~m
depth and no cross-bearing ray forcing. DART and BELLHOP3D
agree to 0.11~dB RMS in full three dimensions
(Table~\ref{tab:accuracy} and Figs.~\ref{fig:fidelity}f
and~\ref{fig:fid3d}a,b). In Nx2D, matching the numerical conventions
reduces the difference between the solvers from 2.15 to 0.11~dB RMS
(Table~\ref{tab:nx2d}).
A separate test showed that matching the
Gaussian minimum-width rule removed a 2.6~dB residual.
Within the last receiver range and with the same tracing box,
two-dimensional BELLHOP and BELLHOP3D in Nx2D mode agree to
$7.6\times10^{-6}$~dB. The reference Nx2D and full three-dimensional
fields differ by 1.00~dB RMS. Since this case has no cross-bearing
ray forcing, that difference reflects how the two formulations
discretize the field.

The Munk bump uses the same sound-speed profile over a 2000~m Gaussian
seamount. The full three-dimensional fields agree to 0.26~dB RMS
and reproduce the shadow behind the bump (Figs.~\ref{fig:fidelity}f
and~\ref{fig:fid3d}c,d). In Nx2D, the result depends on how the seabed
is represented along each bearing. The reference triangulates each
bathymetry cell, placing radial breakpoints at grid lines and cell
diagonals. Bilinear interpolation changes the seabed depth along some
radials by up to 13~m. Matching the discretization reduces the Nx2D
difference from 4.92 to 0.97~dB RMS. Using the reference reflection
rule for verification reduces it further to 0.048~dB RMS. The reference
Nx2D and full three-dimensional fields differ by 4.16~dB RMS
(Table~\ref{tab:nx2d}).

\begin{figure*}[tbp]
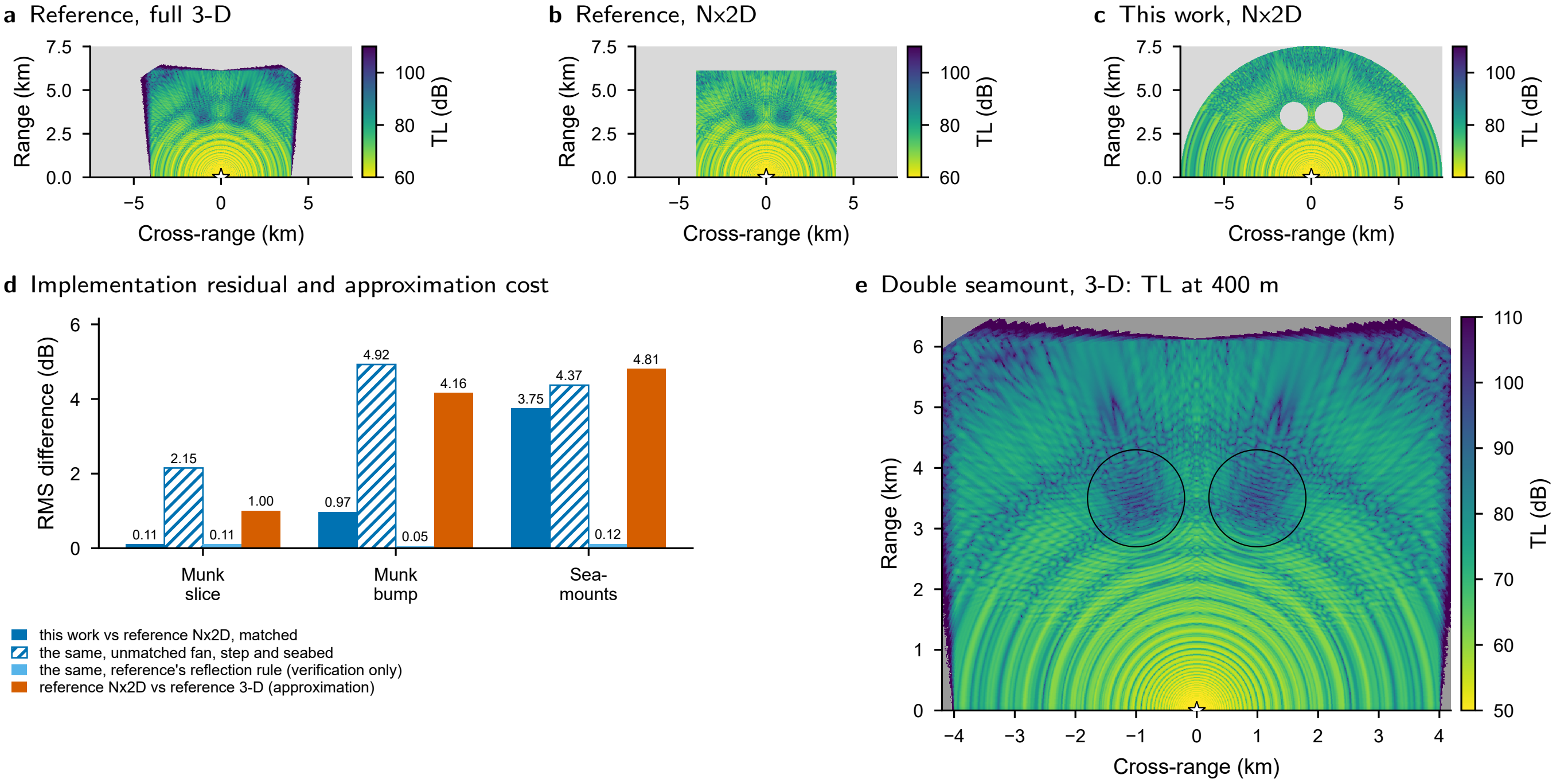

\centering
\pnl{53.5mm}{a}{Reference, full 3-D}{panels/nx2d_dseam_bellhop_3d.pdf}\hfill
\pnl{53.5mm}{b}{Reference, Nx2D}{panels/nx2d_dseam_bellhop_nx2d.pdf}\hfill
\pnl{53.5mm}{c}{This work, Nx2D}{panels/nx2d_dseam_jax.pdf}\\[2mm]
\pnl{81mm}{d}{Implementation residual and approximation cost}{panels/nx2d_stats.pdf}\hfill
\pnl{81mm}{e}{Double seamount, 3-D: TL at 400 m}{panels/fid_dseam_tl.pdf}
\caption{Double-seamount comparison at 100~Hz and 400~m depth.
(a)--(c) Reference full 3-D, reference Nx2D, and DART Nx2D
fields on 721 bearings. White disks mark receivers below the seabed. (d) RMS differences
across cases. (e) DART full 3-D field.}
\label{fig:nx2d}
\end{figure*}

The double-seamount case, in which the two peaks rise to 200~m depth,
separates bearings that cross the relief from those that miss it.
The full three-dimensional fields agree to 0.09~dB RMS
(Table~\ref{tab:accuracy} and Fig.~\ref{fig:fidelity}f).
Along the seamount axis, sampling the triangulated seabed every 20~m
gives an Nx2D difference of 0.70~dB RMS. Using the exact breakpoints,
spaced about 8~m apart, reduces it to 0.03~dB. At matched
discretizations, the Nx2D implementations differ by 0.025~dB RMS on
bearings that miss the relief and by 3.75~dB RMS across the full fan
(Fig.~\ref{fig:nx2d} and Table~\ref{tab:nx2d}).

The larger difference across the fan comes mainly from the reflection
rule on facets that slope across the bearing. The BELLHOP routine projects the seabed normal into the vertical
plane. It computes the dot product with the incident direction before
normalizing the projection. This can change the reflected slowness
magnitude and the grazing angle used in Eq.~(\ref{eq:rayleigh}).
DART uses the unit projected normal throughout the reflection
calculation. Using the reference rule for verification reduces the
difference to 0.12~dB RMS across 721 bearings, with a median difference
of 0.010~dB and 99.9\% of receivers within 1~dB.

The reference Nx2D and full three-dimensional fields differ by
4.81~dB RMS on the double seamount (Table~\ref{tab:nx2d}). Both locate
the two acoustic shadows, while cross-bearing paths in the full
three-dimensional field partly fill the wakes (Fig.~\ref{fig:nx2d}).
Over sloping seabeds, this comparison reflects both the omission of
cross-bearing propagation and the reference reflection rule. The
sensitivity of the coherent field to the launch fan and facet geometry
motivates the use of incoherent and window-averaged observables below.

\begin{table*}[tbp]
\caption{Matched-input differences over the illuminated field.
The median is the median absolute difference. The final column gives
the fraction of illuminated receivers within 1~dB. Receiver counts
describe the comparison grids.}
\label{tab:accuracy}
\begin{ruledtabular}
\begin{tabular}{lrrrrr}
Case & Beams & Receivers & RMS (dB) & Median (dB) & ${\le}1$ dB \\
\hline
Munk, 2-D, hat, 50 Hz          & 999    & 501,501   & 0.51 & 0.023 & 99.4\% \\
Dickins, 2-D, Gaussian, 230 Hz & 4,672  & 601,601   & 1.16 & 0.049 & 92.5\% \\
Munk slice, 3-D, Gaussian      & 25,551 & 1,003,002 & 0.11 & 0.006 & 99.8\% \\
Munk bump, 3-D, hat            & 28,600 & 36,036    & 0.26 & 0.003 & 99.6\% \\
Double seamount, 3-D, Gaussian & 11,300 & 1,082,221 & 0.09 & 0.010 & 99.9\% \\
\end{tabular}
\end{ruledtabular}
\end{table*}

\begin{table*}[tbp]
\caption{Solver comparisons in dB. The reference is BELLHOP3D.
Reference reflection is used for verification only.
Medians are absolute and biases are signed.}
\label{tab:nx2d}
\begin{ruledtabular}

\begin{tabular}{lrrrr}
\multicolumn{5}{c}{(a) DART versus reference Nx2D} \\[2mm]
Case
& \shortstack{Unmatched\\RMS}
& \shortstack{Matched\\RMS}
& \shortstack{Matched\\median}
& \shortstack{Reference reflection\\RMS} \\
\hline
Munk slice (flat) & 2.15 & 0.11 & 0.011 & 0.11 \\
Munk bump         & 4.92 & 0.97 & 0.012 & 0.05 \\
Double seamount   & 4.37 & 3.75 & 0.123 & 0.12 \\
\end{tabular}

\par\vspace{4mm}

\begin{tabular}{lrrr}
\multicolumn{4}{c}{(b) Reference Nx2D versus 3-D and full 3-D agreement} \\[2mm]
Case
& \shortstack{Reference Nx2D--3-D\\bias}
& \shortstack{Reference Nx2D--3-D\\RMS}
& \shortstack{DART--reference 3-D\\RMS} \\
\hline
Munk slice (flat) & $-0.06$ & 1.00 & 0.11 \\
Munk bump         & $-2.18$ & 4.16 & 0.26 \\
Double seamount   & $-1.78$ & 4.81 & 0.09 \\
\end{tabular}

\end{ruledtabular}
\end{table*}
\subsection{Forward computational cost}
\label{sec:speed}

In this subsection, we compare runtimes across the benchmark cases.
Figure~\ref{fig:fidelity}g reports the times of the matched-input
comparisons. The median times below come from repeated runs after
compilation. Munk takes 0.65~s on the GPU compared with 1.14~s for
single-core BELLHOP. Dickins takes 10.8~s compared with 7.6~s.
For full three-dimensional propagation, the GPU solver is 5.7 times
faster on the Munk slice and 10.6 times faster on the double seamount. It is 1.4 times slower on the smaller Munk-bump grid.
We also compare GPU and sixteen-core CPU execution of the same
JAX program. Speedups range from $1.0$ to $10.8\times$ across the
tested forward schedules and gradients, as summarized in
Table~\ref{tab:implspeed}. The CPU uses iterative searches without
scan unrolling or CUDA graphs. For the 241-beam Munk case, the GPU gives speedups
of 1.2 for the forward solve, 1.0 for the differentiable schedule,
and 1.7 for the gradient. The forward speedup increases to 2.0
with 1000 beams. For the 361-beam Dickins case, the speedups are
2.7 for the forward solve and 2.1 for the gradient.
The largest GPU speedup is 10.8 for Dickins with 4600 beams and
a 20~m step, reducing the runtime from 35.5 to 3.28~s. This
configuration takes 8134 steps, compared with 32,529 for the
5~m-step matched-input comparison. The smaller step in that
comparison accounts for its longer GPU runtime of 10.8~s.
These timings depend on the configuration and were measured in double precision on an NVIDIA RTX 2000 Ada Generation Laptop GPU with 8 GB of memory.

\subsection{Verification and cost of the derivatives}
\label{sec:gradients}

\begin{figure*}[tbp]
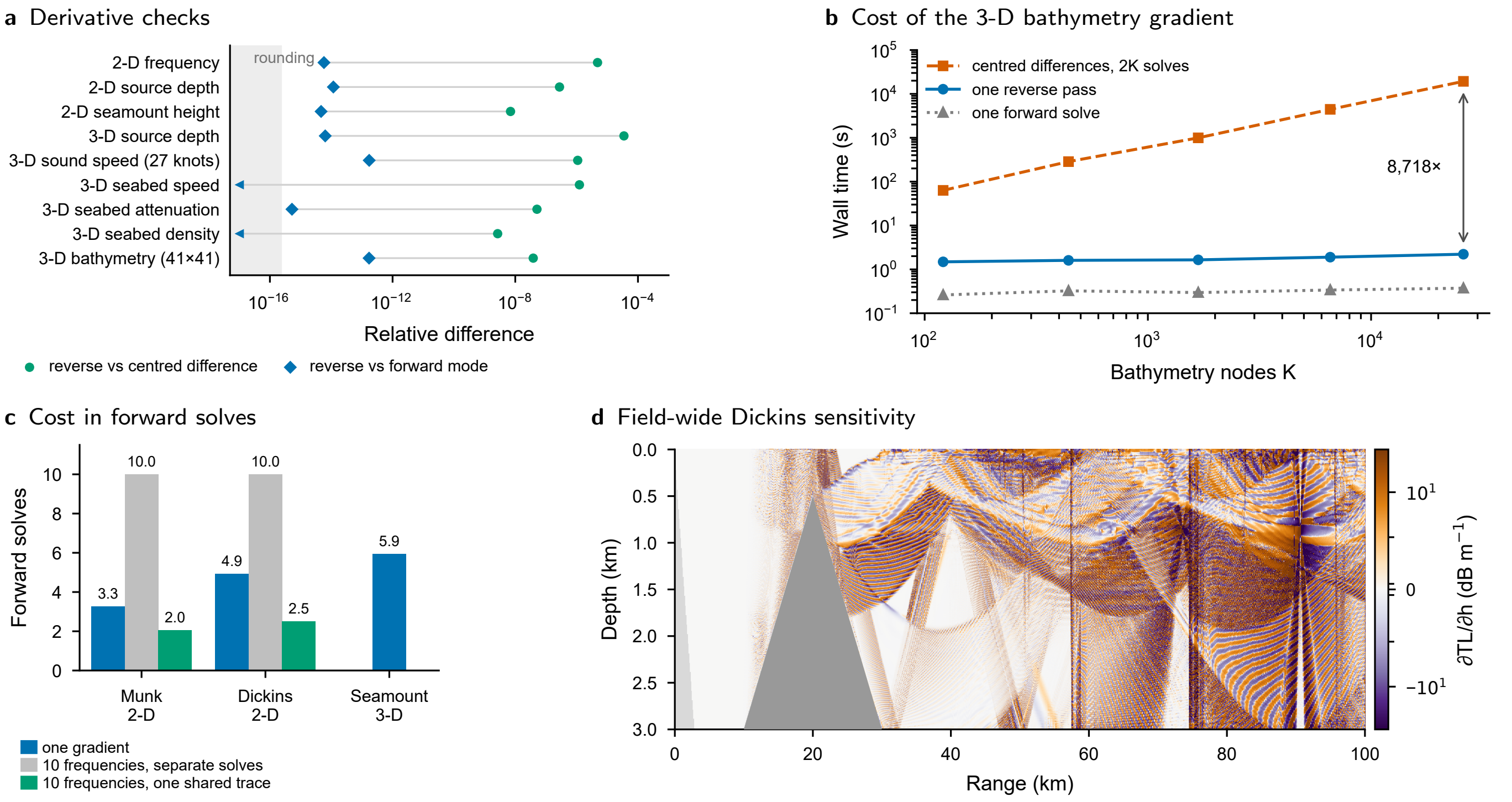

\centering
\pnl{81mm}{a}{Derivative checks}{panels/der_accuracy.pdf}\hfill
\pnl{81mm}{b}{Cost of the 3-D bathymetry gradient}{panels/der_cost.pdf}\\[2mm]
\pnl{53.5mm}{c}{Cost in forward solves}{panels/der_solves.pdf}\hfill
\pnl{109mm}{d}{Field-wide Dickins sensitivity}{panels/der_dickins.pdf}
\caption{(a) Hat-beam derivative checks. (b) Bathymetry-gradient
runtime and estimated finite-difference cost. (c) Gradient and
frequency-sweep costs in forward solves. (d) Dickins TL sensitivity
to seamount height on a symmetric-log scale.}
\label{fig:gradients}
\end{figure*}

In this subsection, we compare forward and reverse differentiation,
check selected derivatives against centred differences, and measure
the cost of computing gradients. We report the relative discrepancy
as $|g_{\mathrm{AD}}-g_{\mathrm{ref}}|/|g_{\mathrm{ref}}|$, where
$g_{\mathrm{ref}}$ is the forward-mode derivative or the centred
difference. For the hat-beam field-energy objectives, forward and
reverse derivatives agree within $1.7\times10^{-13}$ relative.
The relative discrepancies from centred differences range from
$2.7\times10^{-9}$ to $3.5\times10^{-5}$ (Fig.~\ref{fig:gradients}a).
For the Gaussian-beam tests, forward and reverse derivatives agree
within $2.2\times10^{-11}$ relative. The independent checks for the
Gaussian receiver-window kernels and inversion Jacobian are reported
with those results in Secs.~\ref{sec:kernels} and~\ref{sec:inversion}.

We next examine how the cost of a reverse gradient changes with the
number of bathymetry parameters. In the three-dimensional seamount
test, increasing the grid from 121 to 25,921 nodes raises the
gradient time from 1.48 to 2.22~s (Fig.~\ref{fig:gradients}b).
Sequential centred differences require two forward solves per node
and would take about 5.4~h on the largest grid. The reverse gradient
costs the equivalent of 3.3 forward solves for the 999-beam Munk
case, 4.9 for the 4672-beam Dickins case, and 5.9 for the
25,921-node seamount (Fig.~\ref{fig:gradients}c).

Figure~\ref{fig:gradients}c shows the saving from sharing one ray
trace across ten frequencies. The fields agree with those from
separate propagation solves, at a total cost equivalent to 2.0
single-frequency solves for Munk and 2.5 for Dickins. Forward
differentiation also uses the traced field to compute how TL at
every receiver responds to one parameter. In Dickins, one pass
returns 601,601 TL derivatives with respect to seamount height
in 52~s (Fig.~\ref{fig:gradients}d). The derivative is zero at
10.3\% of illuminated receivers and has a median magnitude of
1.06~dB per metre. Small height changes can produce abrupt field
changes, so centred differences do not provide a stable check.
Slope estimates that exclude these jumps support the interpretation
of the map as a local sensitivity.

\subsection{Sensitivity kernels and their linear range}
\label{sec:kernels}

In this subsection, we use sensitivity kernels to identify which
parts of the environment most strongly affect transmission loss (TL).
Each kernel value $K_i$ gives the local change in a chosen TL measure
per unit change in parameter $\theta_i$. For small perturbations,
we estimate the change as
$\Delta\mathrm{TL}\approx\sum_i K_i\Delta\theta_i$.
We compare this prediction with a new propagation calculation to
determine the range over which the derivatives remain useful.

\begin{figure*}[tbp]
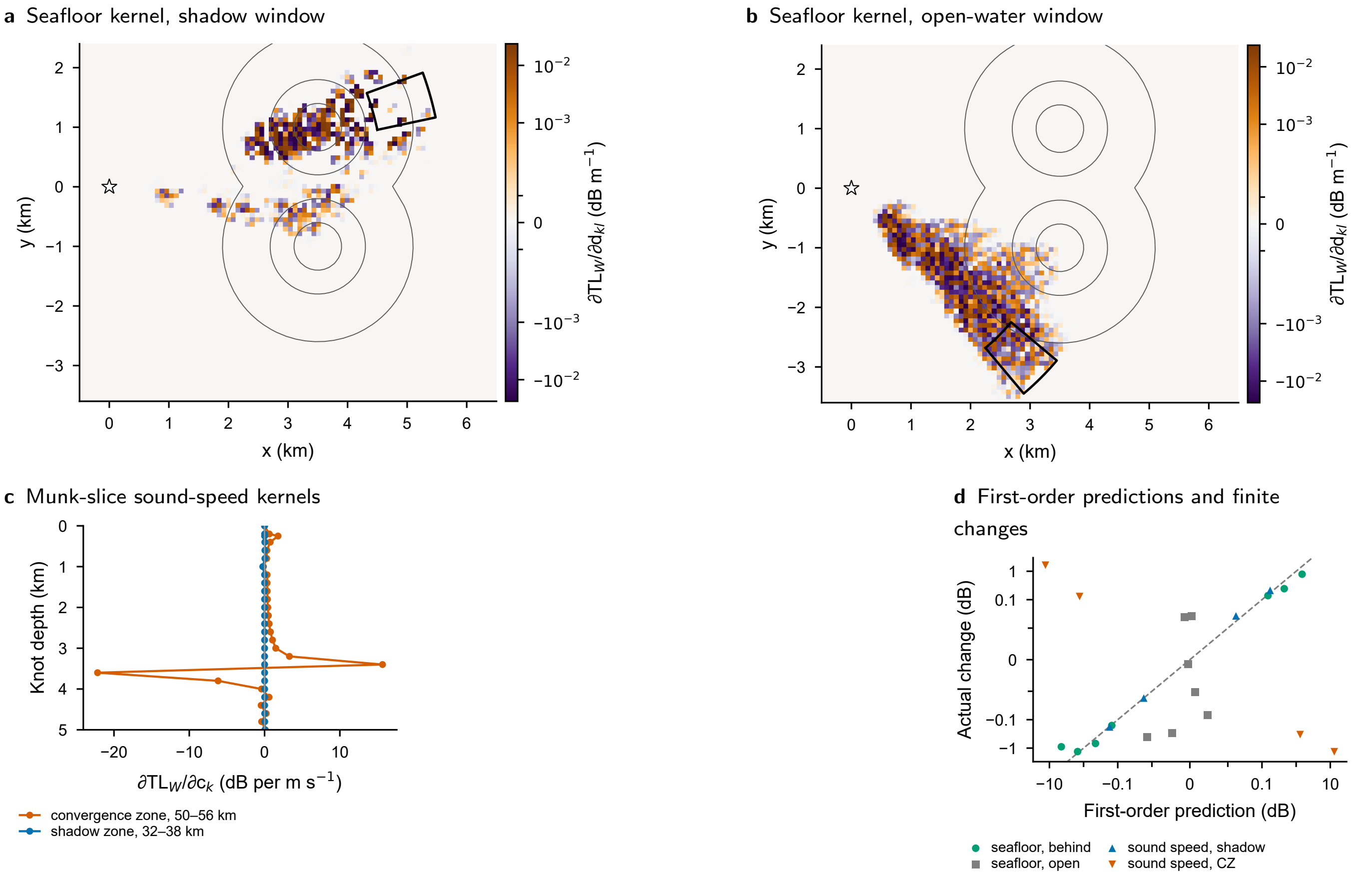

\centering
\pnl{81mm}{a}{Seafloor kernel, shadow window}{panels/der_sea_shadow.pdf}\hfill
\pnl{81mm}{b}{Seafloor kernel, open-water window}{panels/der_sea_open.pdf}\\[2mm]
\pnl{53.5mm}{c}{Munk-slice sound-speed kernels}{panels/der_ssp.pdf}\hfill
\pnl{53.5mm}{d}{First-order predictions and finite changes}{panels/der_linear.pdf}
\caption{Sensitivity kernels and their linear range.
(a,b) Seabed-depth kernels for the outlined windows at 400~m.
The star marks the source. (c) Munk sound-speed knot sensitivities.
(d) First-order predictions and recomputed TL changes.
The dashed line marks equality.}
\label{fig:kernels}
\end{figure*}

We first consider two receiver windows at 400~m depth in the
double-seamount case (Fig.~\ref{fig:kernels}a,b). The shadow window
covers bearings $12^\circ$--$20^\circ$ and ranges 4.6--5.6~km.
The open-water window covers bearings $-50^\circ$ to $-40^\circ$
and ranges 3.5--4.5~km. Each contains $9\times11$ receivers.
For each window, we compute the sensitivity of TL obtained from
window-averaged coherent intensity to 40,000 seabed depths. The kernel
tests use a separate $100\times112$ beam fan, compared with
$100\times113$ for the matched-input benchmark in Table~\ref{tab:setup}.
We divide the kernel fan into 14 azimuthal blocks using
Eq.~(\ref{eq:blocks}), and each kernel takes 30--60~s to compute.

The kernels show which seabed regions influence each window.
Behind the northern seamount, sensitivity is strongest on the
flank facing away from the source and reaches 0.36~dB per metre
of change in node depth. In open water, the kernel follows
bottom-interacting paths and its peak is about one third as large.
For the tested directions of seabed change, the derivatives agree
with centred differences to about $10^{-8}$--$10^{-7}$ relative
at representative step sizes.

These checks establish the local slopes. To test their predictions
for larger changes, we apply a Gaussian seabed perturbation centred
at the peak kernel node and recompute the field
(Fig.~\ref{fig:kernels}d). The depth change is largest at the centre
and decreases smoothly with distance, with a standard deviation of
250~m. Positive perturbations deepen the seabed. In the shadow
window, the first-order prediction agrees within 8\% for $\pm1$~m
perturbations. The discrepancy rises to 12--53\% for amplitudes
of 3--10~m, and the response levels off for a 30~m depression.
In open water, the predicted changes remain below 0.015~dB in
magnitude, while the recomputed changes reach 0.4~dB. The useful
range of a local derivative therefore depends on the receiver
window and the size of the environmental change.

The Munk slice shows how sensitivity also depends on the depth
of a sound-speed change. The convergence-zone kernel is strongest
at 3.4--3.8~km, near the lower ray turning points, while the
preceding shadow zone is about 100 times less sensitive
(Fig.~\ref{fig:kernels}c). Shifting the entire profile gives only
0.15~dB per m~s$^{-1}$. We test finite changes using a Gaussian
sound-speed anomaly with a standard deviation of 300~m. In the
convergence zone, the recomputed TL changes in the opposite direction
to the local prediction (Fig.~\ref{fig:kernels}d). Small perturbations
near rays tangent to profile knots can change which crossings occur.

The double-seamount shadow kernel thus provides both a map of
influential seabed regions and a first-order prediction over the
tested metre-scale changes. The other windows show why the useful
range must be assessed for each observable. We next use the
double-seamount environment to examine parameter recovery from
acoustic observations.

\subsection{Environmental inversion and local uncertainty}
\label{sec:inversion}

In this subsection, we use the solver derivatives to recover
environmental parameters from noisy synthetic BELLHOP3D observations.
We consider the double-seamount case and estimate six parameters:
the relief scale of each seamount, $s_1$ and $s_2$, the surface
sound speed $c_0$ and its depth gradient $g$, and the seabed sound
speed $c_p$ and attenuation $\alpha_p$. The seabed depth is
$d_{\max}-s_1r_1-s_2r_2$, where $r_1$ and $r_2$ describe the
reference relief associated with each peak. The water-column
profile is $c(z)=c_0+gz$, and seabed density is fixed at
1.5~g~cm$^{-3}$. Table~\ref{tab:invparams} lists the priors,
synthetic truth, and parameter bounds.

\begin{table*}[tbp]
\caption{Prior means and standard deviations, synthetic truth,
and inversion bounds.}
\label{tab:invparams}
\begin{ruledtabular}
\begin{tabular}{lrrrrrr}
 & $s_1$ & $s_2$ & $c_0$ (m s$^{-1}$) & $g$ (s$^{-1}$)
 & $c_p$ (m s$^{-1}$) & $\alpha_p$ (dB/$\lambda$) \\
\hline
Prior mean & 1.00 & 1.00 & 1500 & 0 & 1650 & 0.50 \\
Prior s.d. & 0.15 & 0.15 & 5 & 0.02 & 75 & 0.25 \\
Truth & 0.85 & 1.12 & 1494 & 0.017 & 1720 & 0.30 \\
Bounds & 0.3--1.7 & 0.3--1.7 & 1460--1540 & $\pm0.05$
       & 1550--1950 & 0.02--1.5 \\
\end{tabular}
\end{ruledtabular}
\end{table*}

\begin{figure*}[tbp]
\centering
\pnl{53.5mm}{a}{Inversion data at 400 m}{panels/inv_data.pdf}\hfill
\pnl{53.5mm}{b}{Misfit, prior to truth}{panels/inv_line.pdf}\hfill
\pnl{53.5mm}{c}{Damped Gauss--Newton}{panels/inv_conv.pdf}\\[1mm]
\includegraphics{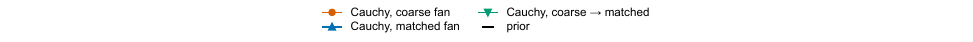}\\[2mm]
\pnl{81mm}{d}{Recovered parameters}{panels/inv_params.pdf}
\caption{Double-seamount inversion.
(a) True TL at 400~m with receivers in white.
(b) Prior-to-truth misfits.
(c) Misfit versus Jacobian evaluations.
(d) Parameter errors in prior units with two local standard
deviations. The error-corrected coarse estimate uses the Gaussian
model-error refit. Black bars mark the prior.}
\label{fig:inversion}
\end{figure*}

The observations are incoherent TL at 100~Hz with 1~dB Gaussian
noise, sampled at 13 bearings over $\pm90^\circ$, three depths
(100, 250, and 400~m), and 29 ranges from 0.5 to 7.5~km.
Illumination and fan screening retain 795 receivers for the coarse
model and 789 for the matched model (Fig.~\ref{fig:inversion}a).
The coarse fan has $100\times113$ beams. The data and matched model
use $200\times226$ beams and a 10~m step. Forward differentiation
computes the six-column Jacobian in 16 and 74~s, respectively.
Centred differences verify the coarse-model relief and seabed
columns at the prior to about $10^{-9}$--$10^{-8}$ relative.
The sound-speed columns have limited or inconsistent agreement.
A coarse-fan bearing follows facet edges, where rounding can
change which facet reflects the ray.

In prior-normalized coordinates,
$\boldsymbol{\theta}=\boldsymbol{\theta}_{\mathrm{prior}}
+\boldsymbol{\sigma}_{\mathrm{prior}}\odot\mathbf z$, we minimize
\begin{equation}
\label{eq:objective}
\begin{aligned}
\Phi(\mathbf z)
&=\sum_i\rho(r_i)+\tfrac12|\mathbf z|^2,
& r_i&=\frac{\mathrm{TL}_i(\boldsymbol{\theta})-d_i}{\sigma_d},\\
\rho(r)
&=\tfrac12c_\rho^2\ln\!\left(1+r^2/c_\rho^2\right),
& w_i&=\left(1+r_i^2/c_\rho^2\right)^{-1}.
\end{aligned}
\end{equation}
Here $\sigma_d=1$~dB and $c_\rho=1$ corresponds to a 1~dB
Cauchy scale. Bounded, damped Gauss--Newton uses the weights
$w_i$ to reduce the influence of large residuals
\cite{levenberg1944method,marquardt1963algorithm}.

The coarse fit reduces the median misfit from 1.12 to 0.71~dB in
four accepted steps, close to the noise-only value of 0.67~dB
(Fig.~\ref{fig:inversion}b,c). Across 163,000 receivers, the median
field error falls from 0.96 to 0.18~dB. Using this estimate to
initialize the matched model reduces the error to 0.03~dB, with
97.8\% of receivers within 1~dB. Five parameters lie within
0.3 prior standard deviations of the truth, while surface sound
speed remains unresolved (Fig.~\ref{fig:inversion}d).

A close field fit can still leave errors in the recovered parameters.
With independent data errors, the local two-standard-deviation
intervals contain the truth for only two parameters in the coarse
fit and four in the coarse-to-matched fit. We therefore account
for differences between the numerical models as well as measurement
noise. We estimate
$\mathbf e=\mathrm{TL}_{\mathrm{BELLHOP3D}}
-\mathrm{TL}_{\mathrm{model}}$ from 48 bound-clipped prior draws.
The mean $\boldsymbol\mu_e$ corrects the predicted levels, while
the covariance $\mathbf C_e$ describes how model errors vary
together across receivers~\cite{kaipio2007statistical,dosso2006data}:
\begin{equation}
\label{eq:modelerror}
\mathbf d=\mathrm{TL}(\boldsymbol{\theta})
+\boldsymbol\mu_e+\boldsymbol\epsilon,
\qquad
\boldsymbol\epsilon\sim
\mathcal N\!\left(\mathbf 0,\sigma_d^2\mathbf I+\mathbf C_e\right).
\end{equation}

Starting from the Cauchy estimate, we fit the parameters again
using this Gaussian error model and the same prior. The corrected
residual $\mathbf r_b$ enters the objective
\begin{equation}
\label{eq:map_b}
\Phi_b(\mathbf z)
=\tfrac12\mathbf r_b^{\mathsf T}\boldsymbol\Sigma_b^{-1}\mathbf r_b
+\tfrac12|\mathbf z|^2,
\qquad
\mathbf r_b=\mathrm{TL}(\boldsymbol{\theta})
+\boldsymbol\mu_e-\mathbf d.
\end{equation}
The combined data covariance $\boldsymbol\Sigma_b$ accounts for
measurement noise and model error. Using the Jacobian at the
new estimate, we approximate the local parameter covariance
as~\cite{tarantola2005inverse}
\begin{equation}
\label{eq:posterior_cov}
\boldsymbol\Sigma_b=\sigma_d^2\mathbf I+\mathbf C_e,
\qquad
\mathbf C_{\mathrm{loc}}
=\left(\mathbf J^{\mathsf T}\boldsymbol\Sigma_b^{-1}\mathbf J
+\mathbf I\right)^{-1},
\end{equation}
where $\mathbf J$ differentiates TL with respect to the
prior-normalized parameters $\mathbf z$.

For the coarse model, the refit accepts two steps. The errors in
$s_2$ and attenuation fall from 0.18 to 0.02 and from 1.12 to
0.10 prior standard deviations, respectively. The local standard
deviations change by factors of 0.91--1.61 relative to the
original Cauchy fit. Five parameters then lie within two local
posterior standard deviations of the truth, while the sound-speed
gradient remains 2.7 standard deviations away.

For the matched model, the discrepancy is only 0.04 dB RMS, and the Gaussian refit accepts no steps. The original Cauchy objective is lower at the truth than at the fitted estimate (195.0 versus 201.8), showing that a better solution exists for that objective. The local uncertainty intervals describe the curvature around the fitted parameters and cannot represent other possible solutions.

The inversion demonstrates how the solver derivatives support environmental estimation from noisy acoustic observations. Using the coarse estimate to initialize the matched model reduces the median field error to 0.03 dB and brings five parameters within 0.3 prior standard deviations of the truth. 
Accounting for model error also improves the coarse-model estimates, with five parameters lying within two local posterior standard deviations. However, surface sound speed remains unresolved in the coarse-to-matched fit, and the sound-speed gradient remains outside this interval after the coarse-model correction. These results show that accurate field recovery can precede complete parameter recovery. Reliable uncertainty estimates also depend on model discrepancy, the information available in the observations, and the solution reached by the optimizer.

\section{\label{sec:6} Conclusions}
DART combines GPU execution with event-consistent differentiation of two- and three-dimensional Gaussian-beam propagation, interface jumps, and sparse receiver-field accumulation. Matched-input comparisons with BELLHOP and BELLHOP3D demonstrate close numerical agreement. The launch-fan and Nx2D tests help identify how implementation choices, numerical resolution, and the restriction to vertical-plane propagation affect the predicted field. Validation against acoustic measurements is still needed to assess accuracy in real environments.

Reverse mode computes a gradient for 25,921 bathymetry nodes at the equivalent cost of 5.9 forward solves. Kernels identify the seabed regions and sound-speed depths controlling the field. Finite-perturbation tests show where local sensitivities remain predictive and where sound-speed knots and facet transitions limit first-order estimates. Synthetic inversion improves field recovery before every environmental parameter is resolved. Accounting for correlated model discrepancy changes the local uncertainty estimates, but does not resolve multimodality or establish calibrated posterior coverage.

Centred-difference checks confirm selected three-dimensional Gaussian-beam kernel derivatives and the relief and seabed columns of the inversion Jacobian at the prior parameters. These checks support using the derivatives to identify influential environmental features and guide parameter estimation. Their computational cost is currently limited by double-precision GPU performance and the memory needed to differentiate boundary interactions. The formulation also assumes fluid seabeds and sound-speed profiles that vary only with depth in three dimensions. Testing has used synthetic observations, so comparisons with hydrophone measurements are needed to assess performance in real environments. Future work will examine smooth environmental representations and extend the formulation to three-dimensional Červený beams and elastic seabeds. These developments would broaden the environmental conditions and applications that the solver can address. The present results provide a basis for environmental inversion and operational acoustic prediction, with finite-perturbation tests establishing where local derivatives can predict environmental changes.

\begin{acknowledgments}
This study is supported by Mitacs, Transport Canada, and Clear Seas through the Quiet-Vessel Initiative (QVI). We thank Dr.~Paul Blomerus and Ms.~Tessa Coulthard for feedback, and Dr.~David Rosen and Dr.~Andrew W.~Trites for advice on marine biology and conservation.
\end{acknowledgments}

\section*{Author Declarations}
\subsection*{Conflict of interest}
The authors have no conflicts to disclose.

\section*{Data Availability}
The data that support the findings of this study are available from the corresponding author upon reasonable request. The numerical examples use prescribed benchmark environments and synthetic acoustic observations.

\section*{References}
\bibliography{diff_beam_refs,companion_refs}

\begin{thebibliography}{50}
\def\enquote#1{``#1,''}
\def\plainquote#1{``#1''}
\expandafter\ifx\csname natexlab\endcsname\relax\def\natexlab#1{#1}\fi
\providecommand{\dourl}[1]{\href{http://#1}{\nolinkurl{#1}}}
\providecommand{\bibinfo}[2]{#2}
\providecommand{\noopsort}[1]{}
\providecommand{\switchargs}[2]{#2#1}
  \def\eatspace #1{#1}

\bibitem[{Ainslie and McColm(1998)}]{ainslie1998simplified}
\bibinfo{author}{Ainslie, M.~A.},  and \bibinfo{author}{McColm, J.~G.} (\textbf{\bibinfo{year}{1998}}). \enquote{\bibinfo{title}{A simplified formula for viscous and chemical absorption in sea water}} \bibinfo{journal}{The Journal of the Acoustical Society of America} \textbf{103}(3), \bibinfo{pages}{1671--1672}, \dodoi{10.1121/1.421258}.

\bibitem[{Baggeroer \emph{et~al.}(1993)Baggeroer, Kuperman, and Mikhalevsky}]{baggeroer1993overview}
\bibinfo{author}{Baggeroer, A.~B.}, \bibinfo{author}{Kuperman, W.~A.},  and \bibinfo{author}{Mikhalevsky, P.~N.} (\textbf{\bibinfo{year}{1993}}). \enquote{\bibinfo{title}{An overview of matched field methods in ocean acoustics}} \bibinfo{journal}{IEEE Journal of Oceanic Engineering} \textbf{18}(4), \bibinfo{pages}{401--424}, \dodoi{10.1109/48.262292}.

\bibitem[{Bezgin \emph{et~al.}(2023)Bezgin, Buhendwa, and Adams}]{bezgin2023jaxfluids}
\bibinfo{author}{Bezgin, D.~A.}, \bibinfo{author}{Buhendwa, A.~B.},  and \bibinfo{author}{Adams, N.~A.} (\textbf{\bibinfo{year}{2023}}). \enquote{\bibinfo{title}{{JAX-Fluids}: A fully-differentiable high-order computational fluid dynamics solver for compressible two-phase flows}} \bibinfo{journal}{Computer Physics Communications} \textbf{282}, \bibinfo{pages}{108527}, \dodoi{10.1016/j.cpc.2022.108527}.

\bibitem[{Bradbury \emph{et~al.}(2018)Bradbury, Frostig, Hawkins, Johnson, Katariya, Leary, Maclaurin, Necula, Paszke, Vander{P}las, Wanderman-{M}ilne, and Zhang}]{jax2018github}
\bibinfo{author}{Bradbury, J.}, \bibinfo{author}{Frostig, R.}, \bibinfo{author}{Hawkins, P.}, \bibinfo{author}{Johnson, M.~J.}, \bibinfo{author}{Katariya, Y.}, \bibinfo{author}{Leary, C.}, \bibinfo{author}{Maclaurin, D.}, \bibinfo{author}{Necula, G.}, \bibinfo{author}{Paszke, A.}, \bibinfo{author}{Vander{P}las, J.}, \bibinfo{author}{Wanderman-{M}ilne, S.},  and \bibinfo{author}{Zhang, Q.} (\textbf{\bibinfo{year}{2018}}). \plainquote{\bibinfo{title}{{JAX}: composable transformations of {Python}+{NumPy} programs}} \bibinfo{howpublished}{\url{https://github.com/jax-ml/jax}}, \bibinfo{note}{software, version 0.11.2}.

\bibitem[{{\v{C}}erven{\'y}(2001)}]{cerveny2001seismic}
\bibinfo{author}{{\v{C}}erven{\'y}, V.} (\textbf{\bibinfo{year}{2001}}). \emph{\bibinfo{title}{Seismic Ray Theory}} (\bibinfo{publisher}{Cambridge University Press}).

\bibitem[{{\v{C}}erven{\'y} \emph{et~al.}(1982){\v{C}}erven{\'y}, Popov, and P{\v{s}}en{\v{c}}{\'\i}k}]{cerveny1982computation}
\bibinfo{author}{{\v{C}}erven{\'y}, V.}, \bibinfo{author}{Popov, M.~M.},  and \bibinfo{author}{P{\v{s}}en{\v{c}}{\'\i}k, I.} (\textbf{\bibinfo{year}{1982}}). \enquote{\bibinfo{title}{Computation of wave fields in inhomogeneous media---{G}aussian beam approach}} \bibinfo{journal}{Geophysical Journal of the Royal Astronomical Society} \textbf{70}(1), \bibinfo{pages}{109--128}, \dodoi{10.1111/j.1365-246X.1982.tb06394.x}.

\bibitem[{Chitre(2023)}]{chitre2023differentiable}
\bibinfo{author}{Chitre, M.} (\textbf{\bibinfo{year}{2023}}). \enquote{\bibinfo{title}{Differentiable ocean acoustic propagation modeling}} in \emph{\bibinfo{booktitle}{{OCEANS} 2023 -- Limerick}}, \bibinfo{publisher}{IEEE}, pp. \bibinfo{pages}{1--8}, \dodoi{10.1109/OCEANSLimerick52467.2023.10244307}.

\bibitem[{Collins(1989)}]{collins1989applications}
\bibinfo{author}{Collins, M.~D.} (\textbf{\bibinfo{year}{1989}}). \enquote{\bibinfo{title}{Applications and time-domain solution of higher-order parabolic equations in underwater acoustics}} \bibinfo{journal}{The Journal of the Acoustical Society of America} \textbf{86}(3), \bibinfo{pages}{1097--1102}, \dodoi{10.1121/1.398101}.

\bibitem[{Dahlen \emph{et~al.}(2000)Dahlen, Hung, and Nolet}]{dahlen2000frechet}
\bibinfo{author}{Dahlen, F.~A.}, \bibinfo{author}{Hung, S.-H.},  and \bibinfo{author}{Nolet, G.} (\textbf{\bibinfo{year}{2000}}). \enquote{\bibinfo{title}{Fr{\'e}chet kernels for finite-frequency traveltimes---{I}. {T}heory}} \bibinfo{journal}{Geophysical Journal International} \textbf{141}(1), \bibinfo{pages}{157--174}, \dodoi{10.1046/j.1365-246X.2000.00070.x}.

\bibitem[{Deo(2025)}]{Deo_2025_thesis}
\bibinfo{author}{Deo, I.~K.} (\textbf{\bibinfo{year}{2025}}). \enquote{\bibinfo{title}{Physics-guided deep learning for dynamical systems: applications to fluid flow and ocean acoustics}} Ph.D. thesis, \bibinfo{school}{University of British Columbia}, \dourl{https://open.library.ubc.ca/collections/ubctheses/24/items/1.0449988}, \dodoi{10.14288/1.0449988}.

\bibitem[{Deo \emph{et~al.}(2023)Deo, Gao, and Jaiman}]{deo2023_3dCOnvPOF}
\bibinfo{author}{Deo, I.~K.}, \bibinfo{author}{Gao, R.},  and \bibinfo{author}{Jaiman, R.} (\textbf{\bibinfo{year}{2023}}). \enquote{\bibinfo{title}{Combined space–time reduced-order model with three-dimensional deep convolution for extrapolating fluid dynamics}} \bibinfo{journal}{Physics of Fluids} \textbf{35}(4), \bibinfo{pages}{043606}, \dourl{https://doi.org/10.1063/5.0145071}, \dodoi{10.1063/5.0145071}.

\bibitem[{Deo and Jaiman(2022)}]{deo2022POF}
\bibinfo{author}{Deo, I.~K.},  and \bibinfo{author}{Jaiman, R.} (\textbf{\bibinfo{year}{2022}}). \enquote{\bibinfo{title}{Predicting waves in fluids with deep neural network}} \bibinfo{journal}{Physics of Fluids} \textbf{34}(6), \bibinfo{pages}{067108}, \dourl{https://doi.org/10.1063/5.0086926}, \dodoi{10.1063/5.0086926}.

\bibitem[{Deo and Jaiman(2026)}]{DEO2026JCP}
\bibinfo{author}{Deo, I.~K.},  and \bibinfo{author}{Jaiman, R.~K.} (\textbf{\bibinfo{year}{2026}}). \enquote{\bibinfo{title}{Predicting wave dynamics using deep learning with multistep integration inspired attention and physics-based loss decomposition}} \bibinfo{journal}{Journal of Computational Physics} \textbf{567}, \bibinfo{pages}{115333}, \dourl{https://www.sciencedirect.com/science/article/pii/S0021999126006820}, \dodoi{10.1016/j.jcp.2026.115333}.

\bibitem[{Deo \emph{et~al.}(2025{\natexlab{a}})Deo, Venkateshwaran, and Jaiman}]{deo2025differentiable}
\bibinfo{author}{Deo, I.~K.}, \bibinfo{author}{Venkateshwaran, A.},  and \bibinfo{author}{Jaiman, R.~K.} (\textbf{\bibinfo{year}{2025}}{\natexlab{a}}). \enquote{\bibinfo{title}{Data-driven differentiable simulations for three-dimensional range-dependent underwater acoustics}} in \emph{\bibinfo{booktitle}{Proceedings of the International Conference on Offshore Mechanics and Arctic Engineering, OMAE2025}}, \bibinfo{organization}{American Society of Mechanical Engineers}, Vol. 4, p. \bibinfo{pages}{V004T08A031}, \dodoi{10.1115/omae2025-157418}.

\bibitem[{Deo \emph{et~al.}(2025{\natexlab{b}})Deo, Venkateshwaran, and Jaiman}]{deo2024continual}
\bibinfo{author}{Deo, I.~K.}, \bibinfo{author}{Venkateshwaran, A.},  and \bibinfo{author}{Jaiman, R.~K.} (\textbf{\bibinfo{year}{2025}}{\natexlab{b}}). \enquote{\bibinfo{title}{Predicting transmission loss in underwater acoustics using continual learning with range-dependent conditional convolutional neural networks}} \bibinfo{journal}{The Journal of the Acoustical Society of America} \textbf{157}(5), \bibinfo{pages}{3930--3945}, \dodoi{10.1121/10.0036773}.

\bibitem[{Deo \emph{et~al.}(2026)Deo, Venkateshwaran, and Jaiman}]{deo2025data}
\bibinfo{author}{Deo, I.~K.}, \bibinfo{author}{Venkateshwaran, A.},  and \bibinfo{author}{Jaiman, R.~K.} (\textbf{\bibinfo{year}{2026}}). \enquote{\bibinfo{title}{A physics-guided probabilistic surrogate modeling framework for digital twins of underwater radiated noise}} \bibinfo{journal}{Ocean Engineering} \textbf{358}, \bibinfo{pages}{125461}, \dodoi{10.1016/j.oceaneng.2026.125461}.

\bibitem[{Dosso(2002)}]{dosso2002quantifying}
\bibinfo{author}{Dosso, S.~E.} (\textbf{\bibinfo{year}{2002}}). \enquote{\bibinfo{title}{Quantifying uncertainty in geoacoustic inversion. {I}. {A} fast {G}ibbs sampler approach}} \bibinfo{journal}{The Journal of the Acoustical Society of America} \textbf{111}(1), \bibinfo{pages}{129--142}, \dodoi{10.1121/1.1419086}.

\bibitem[{Dosso \emph{et~al.}(2006)Dosso, Nielsen, and Wilmut}]{dosso2006data}
\bibinfo{author}{Dosso, S.~E.}, \bibinfo{author}{Nielsen, P.~L.},  and \bibinfo{author}{Wilmut, M.~J.} (\textbf{\bibinfo{year}{2006}}). \enquote{\bibinfo{title}{Data error covariance in matched-field geoacoustic inversion}} \bibinfo{journal}{The Journal of the Acoustical Society of America} \textbf{119}(1), \bibinfo{pages}{208--219}, \dodoi{10.1121/1.2139625}.

\bibitem[{Duarte \emph{et~al.}(2021)Duarte, Chapuis, Collin, Costa, Devassy, Eguiluz, Erbe, Gordon, Halpern, Harding, Havlik, Meekan, Merchant, Miksis-Olds, Parsons, Predragovic, Radford, Radford, Simpson, Slabbekoorn, Staaterman, Van~Opzeeland, Winderen, Zhang, and Juanes}]{duarte2021soundscape}
\bibinfo{author}{Duarte, C.~M.}, \bibinfo{author}{Chapuis, L.}, \bibinfo{author}{Collin, S.~P.}, \bibinfo{author}{Costa, D.~P.}, \bibinfo{author}{Devassy, R.~P.}, \bibinfo{author}{Eguiluz, V.~M.}, \bibinfo{author}{Erbe, C.}, \bibinfo{author}{Gordon, T. A.~C.}, \bibinfo{author}{Halpern, B.~S.}, \bibinfo{author}{Harding, H.~R.}, \bibinfo{author}{Havlik, M.~N.}, \bibinfo{author}{Meekan, M.}, \bibinfo{author}{Merchant, N.~D.}, \bibinfo{author}{Miksis-Olds, J.~L.}, \bibinfo{author}{Parsons, M.}, \bibinfo{author}{Predragovic, M.}, \bibinfo{author}{Radford, A.~N.}, \bibinfo{author}{Radford, C.~A.}, \bibinfo{author}{Simpson, S.~D.}, \bibinfo{author}{Slabbekoorn, H.}, \bibinfo{author}{Staaterman, E.}, \bibinfo{author}{Van~Opzeeland, I.~C.}, \bibinfo{author}{Winderen, J.}, \bibinfo{author}{Zhang, X.},  and \bibinfo{author}{Juanes, F.} (\textbf{\bibinfo{year}{2021}}). \enquote{\bibinfo{title}{The soundscape of the {Anthropocene} ocean}} \bibinfo{journal}{Science} \textbf{371}(6529), \bibinfo{pages}{eaba4658},
  \dodoi{10.1126/science.aba4658}.

\bibitem[{Erbe \emph{et~al.}(2019)Erbe, Marley, Schoeman, Smith, Trigg, and Embling}]{erbe2019effects}
\bibinfo{author}{Erbe, C.}, \bibinfo{author}{Marley, S.~A.}, \bibinfo{author}{Schoeman, R.~P.}, \bibinfo{author}{Smith, J.~N.}, \bibinfo{author}{Trigg, L.~E.},  and \bibinfo{author}{Embling, C.~B.} (\textbf{\bibinfo{year}{2019}}). \enquote{\bibinfo{title}{The effects of ship noise on marine mammals---a review}} \bibinfo{journal}{Frontiers in Marine Science} \textbf{6}, \bibinfo{pages}{606}, \dodoi{10.3389/fmars.2019.00606}.

\bibitem[{Francois and Garrison(1982)}]{francois1982sound}
\bibinfo{author}{Francois, R.~E.},  and \bibinfo{author}{Garrison, G.~R.} (\textbf{\bibinfo{year}{1982}}). \enquote{\bibinfo{title}{Sound absorption based on ocean measurements. {Part II}: Boric acid contribution and equation for total absorption}} \bibinfo{journal}{The Journal of the Acoustical Society of America} \textbf{72}(6), \bibinfo{pages}{1879--1890}, \dodoi{10.1121/1.388673}.

\bibitem[{Griewank and Walther(2008)}]{griewank2008evaluating}
\bibinfo{author}{Griewank, A.},  and \bibinfo{author}{Walther, A.} (\textbf{\bibinfo{year}{2008}}). \emph{\bibinfo{title}{Evaluating Derivatives: Principles and Techniques of Algorithmic Differentiation}}, \bibinfo{edition}{2nd} ed. (\bibinfo{publisher}{Society for Industrial and Applied Mathematics}).

\bibitem[{Hursky \emph{et~al.}(2004)Hursky, Porter, Cornuelle, Hodgkiss, and Kuperman}]{hursky2004adjoint}
\bibinfo{author}{Hursky, P.}, \bibinfo{author}{Porter, M.~B.}, \bibinfo{author}{Cornuelle, B.~D.}, \bibinfo{author}{Hodgkiss, W.~S.},  and \bibinfo{author}{Kuperman, W.~A.} (\textbf{\bibinfo{year}{2004}}). \enquote{\bibinfo{title}{Adjoint modeling for acoustic inversion}} \bibinfo{journal}{The Journal of the Acoustical Society of America} \textbf{115}(2), \bibinfo{pages}{607--619}, \dodoi{10.1121/1.1636760}.

\bibitem[{{International Maritime Organization}(2024)}]{imo2024urn}
\bibinfo{author}{{International Maritime Organization}} (\textbf{\bibinfo{year}{2024}}). \enquote{\bibinfo{title}{Revised guidelines for the reduction of underwater radiated noise from shipping to address adverse impacts on marine life}} \bibinfo{type}{Circular} \bibinfo{number}{MEPC.1/Circ.906/Rev.1}, \dourl{https://wwwcdn.imo.org/localresources/en/MediaCentre/Documents/MEPC.1-Circ.906-Rev.1\%20-\%20Revised\%20Guidelines\%20For\%20The\%20Reduction\%20Of\%20Underwater\%20Radiated.pdf}, \bibinfo{note}{issued 15 November 2024}.

\bibitem[{Jensen \emph{et~al.}(2011)Jensen, Kuperman, Porter, and Schmidt}]{jensen2011computational}
\bibinfo{author}{Jensen, F.~B.}, \bibinfo{author}{Kuperman, W.~A.}, \bibinfo{author}{Porter, M.~B.},  and \bibinfo{author}{Schmidt, H.} (\textbf{\bibinfo{year}{2011}}). \emph{\bibinfo{title}{Computational Ocean Acoustics}}, \bibinfo{edition}{2nd} ed. (\bibinfo{publisher}{Springer}).

\bibitem[{Kaipio and Somersalo(2007)}]{kaipio2007statistical}
\bibinfo{author}{Kaipio, J.},  and \bibinfo{author}{Somersalo, E.} (\textbf{\bibinfo{year}{2007}}). \enquote{\bibinfo{title}{Statistical inverse problems: discretization, model reduction and inverse crimes}} \bibinfo{journal}{Journal of Computational and Applied Mathematics} \textbf{198}(2), \bibinfo{pages}{493--504}, \dodoi{10.1016/j.cam.2005.09.027}.

\bibitem[{Kirk and Hwu(2017)}]{kirk2016programming}
\bibinfo{author}{Kirk, D.~B.},  and \bibinfo{author}{Hwu, W.-m.~W.} (\textbf{\bibinfo{year}{2017}}). \emph{\bibinfo{title}{Programming Massively Parallel Processors: A Hands-on Approach}}, \bibinfo{edition}{3rd} ed. (\bibinfo{publisher}{Morgan Kaufmann}).

\bibitem[{Levenberg(1944)}]{levenberg1944method}
\bibinfo{author}{Levenberg, K.} (\textbf{\bibinfo{year}{1944}}). \enquote{\bibinfo{title}{A method for the solution of certain non-linear problems in least squares}} \bibinfo{journal}{Quarterly of Applied Mathematics} \textbf{2}(2), \bibinfo{pages}{164--168}, \dodoi{10.1090/qam/10666}.

\bibitem[{Lytaev(2024)}]{lytaev2024automatically}
\bibinfo{author}{Lytaev, M.} (\textbf{\bibinfo{year}{2024}}). \enquote{\bibinfo{title}{Automatically differentiable higher-order parabolic equation for real-time underwater sound speed profile sensing}} \bibinfo{journal}{Journal of Marine Science and Engineering} \textbf{12}(11), \bibinfo{pages}{1925}, \dodoi{10.3390/jmse12111925}.

\bibitem[{Mallik \emph{et~al.}(2022)Mallik, Jaiman, and Jelovica}]{mallik2022predicting}
\bibinfo{author}{Mallik, W.}, \bibinfo{author}{Jaiman, R.~K.},  and \bibinfo{author}{Jelovica, J.} (\textbf{\bibinfo{year}{2022}}). \enquote{\bibinfo{title}{Predicting transmission loss in underwater acoustics using convolutional recurrent autoencoder network}} \bibinfo{journal}{The Journal of the Acoustical Society of America} \textbf{152}(3), \bibinfo{pages}{1627--1638}, \dodoi{10.1121/10.0013894}.

\bibitem[{Marquardt(1963)}]{marquardt1963algorithm}
\bibinfo{author}{Marquardt, D.~W.} (\textbf{\bibinfo{year}{1963}}). \enquote{\bibinfo{title}{An algorithm for least-squares estimation of nonlinear parameters}} \bibinfo{journal}{Journal of the Society for Industrial and Applied Mathematics} \textbf{11}(2), \bibinfo{pages}{431--441}, \dodoi{10.1137/0111030}.

\bibitem[{Munk \emph{et~al.}(1995)Munk, Worcester, and Wunsch}]{munk1995ocean}
\bibinfo{author}{Munk, W.}, \bibinfo{author}{Worcester, P.},  and \bibinfo{author}{Wunsch, C.} (\textbf{\bibinfo{year}{1995}}). \emph{\bibinfo{title}{Ocean Acoustic Tomography}} (\bibinfo{publisher}{Cambridge University Press}).

\bibitem[{Pisha \emph{et~al.}(2023)Pisha, Snider, Jackson, and Jaffe}]{pisha2023bellhopcuda}
\bibinfo{author}{Pisha, L.~A.}, \bibinfo{author}{Snider, J.}, \bibinfo{author}{Jackson, K.},  and \bibinfo{author}{Jaffe, J.~S.} (\textbf{\bibinfo{year}{2023}}). \enquote{\bibinfo{title}{Introducing bellhopcxx/bellhopcuda: Modern, parallel {BELLHOP(3D)}}} \bibinfo{journal}{The Journal of the Acoustical Society of America} \textbf{153}(3\_Supplement), \bibinfo{pages}{A218}, \dodoi{10.1121/10.0018709}.

\bibitem[{Plessix(2006)}]{plessix2006review}
\bibinfo{author}{Plessix, R.-E.} (\textbf{\bibinfo{year}{2006}}). \enquote{\bibinfo{title}{A review of the adjoint-state method for computing the gradient of a functional with geophysical applications}} \bibinfo{journal}{Geophysical Journal International} \textbf{167}(2), \bibinfo{pages}{495--503}, \dodoi{10.1111/j.1365-246x.2006.02978.x}.

\bibitem[{Porter(2011)}]{porter2011bellhop}
\bibinfo{author}{Porter, M.~B.} (\textbf{\bibinfo{year}{2011}}). \enquote{\bibinfo{title}{The {BELLHOP} manual and user's guide: Preliminary draft}} \bibinfo{type}{Technical Report}, \dourl{https://oalib-acoustics.org/website_resources/Rays/HLS-2010-1.pdf}.

\bibitem[{Porter(2019)}]{porter2019beam}
\bibinfo{author}{Porter, M.~B.} (\textbf{\bibinfo{year}{2019}}). \enquote{\bibinfo{title}{Beam tracing for two- and three-dimensional problems in ocean acoustics}} \bibinfo{journal}{The Journal of the Acoustical Society of America} \textbf{146}(3), \bibinfo{pages}{2016--2029}, \dodoi{10.1121/1.5125262}.

\bibitem[{Porter(2026)}]{acoustics_toolbox}
\bibinfo{author}{Porter, M.~B.} (\textbf{\bibinfo{year}{2026}}). \plainquote{\bibinfo{title}{Acoustics toolbox}} \bibinfo{howpublished}{\url{http://oalib.hlsresearch.com/AcousticsToolbox/}}, \bibinfo{note}{release of 29 August 2026}.

\bibitem[{Porter and Bucker(1987)}]{porter1987gaussian}
\bibinfo{author}{Porter, M.~B.},  and \bibinfo{author}{Bucker, H.~P.} (\textbf{\bibinfo{year}{1987}}). \enquote{\bibinfo{title}{Gaussian beam tracing for computing ocean acoustic fields}} \bibinfo{journal}{The Journal of the Acoustical Society of America} \textbf{82}(4), \bibinfo{pages}{1349--1359}, \dodoi{10.1121/1.395269}.

\bibitem[{Porter and Reiss(1984)}]{porter1984normal}
\bibinfo{author}{Porter, M.~B.},  and \bibinfo{author}{Reiss, E.~L.} (\textbf{\bibinfo{year}{1984}}). \enquote{\bibinfo{title}{A numerical method for ocean-acoustic normal modes}} \bibinfo{journal}{The Journal of the Acoustical Society of America} \textbf{76}(1), \bibinfo{pages}{244--252}, \dodoi{10.1121/1.391101}.

\bibitem[{Richardson \emph{et~al.}(1995)Richardson, Greene, Malme, and Thomson}]{richardson1995marine}
\bibinfo{author}{Richardson, W.~J.}, \bibinfo{author}{Greene, Jr., C.~R.}, \bibinfo{author}{Malme, C.~I.},  and \bibinfo{author}{Thomson, D.~H.} (\textbf{\bibinfo{year}{1995}}). \emph{\bibinfo{title}{Marine Mammals and Noise}} (\bibinfo{publisher}{Academic Press}).

\bibitem[{Skarsoulis and Cornuelle(2004)}]{skarsoulis2004travel}
\bibinfo{author}{Skarsoulis, E.~K.},  and \bibinfo{author}{Cornuelle, B.~D.} (\textbf{\bibinfo{year}{2004}}). \enquote{\bibinfo{title}{Travel-time sensitivity kernels in ocean acoustic tomography}} \bibinfo{journal}{The Journal of the Acoustical Society of America} \textbf{116}(1), \bibinfo{pages}{227--238}, \dodoi{10.1121/1.1753292}.

\bibitem[{Stanziola \emph{et~al.}(2023)Stanziola, Arridge, Cox, and Treeby}]{stanziola2023jwave}
\bibinfo{author}{Stanziola, A.}, \bibinfo{author}{Arridge, S.~R.}, \bibinfo{author}{Cox, B.~T.},  and \bibinfo{author}{Treeby, B.~E.} (\textbf{\bibinfo{year}{2023}}). \enquote{\bibinfo{title}{j-{W}ave: An open-source differentiable wave simulator}} \bibinfo{journal}{SoftwareX} \textbf{22}, \bibinfo{pages}{101338}, \dodoi{10.1016/j.softx.2023.101338}.

\bibitem[{Tappert(1977)}]{tappert1977parabolic}
\bibinfo{author}{Tappert, F.~D.} (\textbf{\bibinfo{year}{1977}}). \enquote{\bibinfo{title}{The parabolic approximation method}} in \emph{\bibinfo{booktitle}{Wave Propagation and Underwater Acoustics}}, edited by \bibinfo{editor}{J.~B. Keller} and \bibinfo{editor}{J.~S. Papadakis}, \bibinfo{volume}{\textbf{70}} of \emph{\bibinfo{series}{Lecture Notes in Physics}} (\bibinfo{publisher}{Springer-Verlag}), pp. \bibinfo{pages}{224--287}, \dodoi{10.1007/3-540-08527-0_5}.

\bibitem[{Tarantola(2005)}]{tarantola2005inverse}
\bibinfo{author}{Tarantola, A.} (\textbf{\bibinfo{year}{2005}}). \emph{\bibinfo{title}{Inverse Problem Theory and Methods for Model Parameter Estimation}} (\bibinfo{publisher}{Society for Industrial and Applied Mathematics}).

\bibitem[{Teh \emph{et~al.}(2022)Teh, O'Toole, and Gkioulekas}]{teh2022adjoint}
\bibinfo{author}{Teh, A.}, \bibinfo{author}{O'Toole, M.},  and \bibinfo{author}{Gkioulekas, I.} (\textbf{\bibinfo{year}{2022}}). \enquote{\bibinfo{title}{Adjoint nonlinear ray tracing}} \bibinfo{journal}{ACM Transactions on Graphics} \textbf{41}(4), \bibinfo{pages}{126:1--126:13}, \dodoi{10.1145/3528223.3530077}.

\bibitem[{Thorp(1967)}]{thorp1967analytic}
\bibinfo{author}{Thorp, W.~H.} (\textbf{\bibinfo{year}{1967}}). \enquote{\bibinfo{title}{Analytic description of the low-frequency attenuation coefficient}} \bibinfo{journal}{The Journal of the Acoustical Society of America} \textbf{42}(1), \bibinfo{pages}{270}, \dodoi{10.1121/1.1910566}.

\bibitem[{Tromp \emph{et~al.}(2005)Tromp, Tape, and Liu}]{tromp2005seismic}
\bibinfo{author}{Tromp, J.}, \bibinfo{author}{Tape, C.},  and \bibinfo{author}{Liu, Q.} (\textbf{\bibinfo{year}{2005}}). \enquote{\bibinfo{title}{Seismic tomography, adjoint methods, time reversal and banana-doughnut kernels}} \bibinfo{journal}{Geophysical Journal International} \textbf{160}(1), \bibinfo{pages}{195--216}, \dodoi{10.1111/j.1365-246X.2004.02453.x}.

\bibitem[{Vardi \emph{et~al.}(2026)Vardi, Averbuch, and Leonard}]{vardi2026sensitivity}
\bibinfo{author}{Vardi, A.}, \bibinfo{author}{Averbuch, G.},  and \bibinfo{author}{Leonard, J.~J.} (\textbf{\bibinfo{year}{2026}}). \enquote{\bibinfo{title}{Sensitivity analysis of normal mode algorithms using automatic differentiation}} \bibinfo{journal}{The Journal of the Acoustical Society of America} \textbf{159}(5), \bibinfo{pages}{4576--4589}, \dodoi{10.1121/10.0043870}.

\bibitem[{Venkateshwaran \emph{et~al.}(2026)Venkateshwaran, Deo, and Jaiman}]{venkateshwaran2025mute}
\bibinfo{author}{Venkateshwaran, A.}, \bibinfo{author}{Deo, I.~K.},  and \bibinfo{author}{Jaiman, R.~K.} (\textbf{\bibinfo{year}{2026}}). \enquote{\bibinfo{title}{{MUTE-DSS}: A digital-twin-based decision support system for minimizing underwater radiated noise in ship voyage planning}} \bibinfo{journal}{Ocean Engineering} \textbf{343}, \bibinfo{pages}{123343}, \dodoi{10.1016/j.oceaneng.2025.123343}.

\bibitem[{Venkateshwaran \emph{et~al.}(2024)Venkateshwaran, Deo, Jelovica, and Jaiman}]{venkateshwaran2024multi}
\bibinfo{author}{Venkateshwaran, A.}, \bibinfo{author}{Deo, I.~K.}, \bibinfo{author}{Jelovica, J.},  and \bibinfo{author}{Jaiman, R.~K.} (\textbf{\bibinfo{year}{2024}}). \enquote{\bibinfo{title}{A multi-objective optimization framework for reducing the impact of ship noise on marine mammals}} \bibinfo{journal}{Ocean Engineering} \textbf{310}, \bibinfo{pages}{118687}, \dodoi{10.1016/j.oceaneng.2024.118687}.

\end{thebibliography}
\end{document}